\documentclass[useAMS,usenatbib]{mn2e}  %draft
\usepackage{subcaption}
\usepackage{amssymb}
\usepackage{ifdraft}
\usepackage[dvipsnames]{xcolor}
\usepackage{ifpdf}
\usepackage[a4paper, total={6.5in, 10in}]{geometry}

\ifpdf
\usepackage[pdftex]{graphicx}
\else
\usepackage{graphicx}
\fi

\ifpdf
\DeclareGraphicsExtensions{.pdf, .jpg, .tif, .png}
\else
\DeclareGraphicsExtensions{.eps, .jpg}
\fi

\newcommand{\specialcell}[2][c]{%
  \begin{tabular}[#1]{@{}c@{}}#2\end{tabular}}

\title[Chameleon $f(R)$ gravity at the cluster scale]{Chameleon $f(R)$ gravity on the Virgo cluster scale}
\author[C. Corbett Moran, R. Teyssier, B. Li]{C. Corbett Moran$^{1}$\thanks{E-mail:
corbett@physik.uzh.ch}, R. Teyssier$^{1}$, B. Li$^2$\\
$^{1}$Institute for Theoretical Physics, University of Zurich, Winterthurerstrasse 190, CH-8057, Z\"urich Switzerland\\
$^{2}$Institute for Computational Cosmology, Department of Physics, Durham University, South Road,\\
Durham DH1 3LE, United Kingdom}
\begin{document}
\graphicspath{{./PublishFigures/LowRes/}}

\date{Received 2014 August 15}

\pagerange{1--24} \pubyear{2014}

\maketitle

\label{firstpage}

\begin{abstract}
    Models of modified gravity offer promising alternatives to the concordance $\Lambda$CDM cosmology to explain the late-time acceleration of the universe. A popular such model is $f(R)$ gravity, in which the Ricci scalar in the Einstein-Hilbert action is replaced by a general function of it. We study the $f(R)$ model of Hu \& Sawicki (2007), which recovers standard General Relativity in high density regimes, while reproducing the desired late-time acceleration at cosmological scales. We run a suite of high resolution zoom simulations using the \texttt{ECOSMOG} code to examine the effect of $f(R)$ gravity on the properties of a halo that is analogous to the Virgo cluster. We show that the velocity dispersion profiles can potentially discriminate between $f(R)$ models and $\Lambda$CDM, and provide complementary analysis of lensing signal profiles to explore the possibility to further distinguish the different $f(R)$ models. Our results confirm the techniques explored by Cabre et al. (2012) to quantify the effect of environment in the behavior of $f(R)$ gravity, and we extend them to study halo satellites at various redshifts. We find that the modified gravity effects in our models are most observable at low redshifts, and that effects are generally stronger for satellites far from the center of the main halo. We show that the screening properties of halo satellites trace very well that of dark matter particles, which means that low-resolution simulations in which subhalos are not very well resolved can in principle be used to study satellite properties. We discuss observables, particularly for halo satellites, that can potentially be used to constrain the observational viability of $f(R)$ gravity.	
\end{abstract}

\section{Introduction}
\label{sec:modifiedgravity}
Modern cosmology strives to explain the late time acceleration of the universe \citep{Riess:1998hp}. The commonly accepted candidate is a positive cosmological constant: this is the famed $\Lambda$ in the concordance model of cosmology $\Lambda$CDM. However this leaves several unresolved questions, as its value to match cosmological constraints mismatches with predictions from quantum field theory by more than 100 orders of magnitude \cite{1992ARA&A..30..499C}. 

These unresolved questions have motived the proposal of alternative explanations of the accelerated expansion of the universe. Such alternative models to $\Lambda$CDM can be divided into two groups generally, those that introduce new matter species or dynamical fields (reviewed in \cite{Copeland:2006vf}), essentially changing our understanding of the ingredients of stress-energy side of Einstein's equations, commonly known as dark energy models and model candidates which incorporate modifications to the geometrical portion of Einstein's equations (reviewed in \cite{deFelice:2010el}). We focus this work on one of the most well-studied modified gravity models belonging to the second group, $f(R)$ gravity as reviewed in \cite{Sotiriou:2010hr} in which the Einstein-Hilbert action becomes a general function of the Ricci-scalar.

General relativity (GR) has been confirmed with high accuracy locally, thus such alternative models are highly constrained by local tests \citep{Will:2006iy,Bertotti:2003by,Hoyle:2004vv,Lyne:2004jm,Adelberger:2003zx}. In contrast to dark energy, which functions mainly to modify the cosmic expansion history, alternative gravity models modifying the metric side of Einstein's equations predict a different force law between particles and change structure formation directly. This change is in principle observable, and can be used to discriminate between the two scenarios.

Considering the modification to standard gravity as an effective fifth force, to match local observations, such a fifth force must be suppressed locally to have very weak strength and/or very short (sub-millimeter) range \citep{Li:2012by}. In the case of a fifth force being mediated by a scalar degree of freedom there have been an array of proposals of screening mechanisms to achieve this.

These screening models can themselves be divided into two classes generally. The first class includes works which invoke nonlinear kinetic terms such as the Galileon \citep{Nicolis:2008in,Deffayet:2009wt} using the Vainshtein mechanism \citep{Vainshtein:1972sx} to reduce the fifth force in areas of high density to remain within experimental constraints. The second class of models include Chameleon \citep{Khoury:2003rn}, dilaton \citep{2010PhRvD..82f3519B}, symmetron \citep{Hinterbichler:2010es}, and others (reviewed in  \cite{deFelice:2010el}) and screens the fifth force in dense environments due to nonlinearities of the scalar potential and/or its coupling to matter. We consider a physically plausible $f(R)$ model which incorporates a chameleon type screening mechanism of this second class, the Hu-Sawicki model \citep{Hu:2007tv}. 

In particular, we perform the highest resolution N-Body simulation in $f(R)$ gravity in the Hu-Sawicki model to date. With this resolution we are able to study the properties of an individual halo, with a Virgo analogue mass, and its associated satellites in detail. In Section \ref{sec:theory} we briefly review $f(R)$ gravity models, the chameleon mechanism, the Hu-Sawicki model in particular and past work in simulations of $f(R)$ gravity. In Section \ref{sec:simulation} we detail our numerical methods and suite of simulations run. In Section \ref{sec:visual} we present a qualitative overview of the results of our suite of simulations. In Section \ref{sec:profiles} we present and analyze the line of sight velocity dispersion and surface density distribution profiles and their evolution. In Section \ref{sec:lensing} we present basic lensing theory, observational constraints and our results as to the lensing signal profile and its evolution. In Section \ref{sec:physical} we analyze the fifth force versus the standard gravity forces as a function of environment, focusing on the satellite population. Finally in Section \ref{sec:conclusions} we present our conclusions.

\section{$f(R)$ Gravity Theory, Chameleon Mechanism, and non-linear simulation methods}
\label{sec:theory}
We focus this work on $f(R)$ gravity as reviewed in \cite{Sotiriou:2010hr} in which the Einstein-Hilbert action becomes a general function of the Ricci-scalar: 
\begin{equation}
	S=\int d^4x \sqrt{-g} \left[ \frac{R+f(R)}{16 \pi G} + \mathcal{L}_M \right]
\end{equation}
Here, $\mathcal{L}_M$ is the Lagrangian density for matter fields including radiation, baryons and cold dark matter, and $G$ is the Newtonian gravitational constant. Taking the variation of this with respect to the metric yields the modified Einstein equations, and we introduce $f_R \equiv \frac{df(R)}{dR}$ which corresponds to an extra scalar degree of freedom, the \emph{scalaron}. Equations governing the perturbation of this \emph{scalaron} are, namely its equation of motion
\begin{equation}
	\label{eq:fr}
	\nabla^2 f_R = -\frac{a}{3}\left[ \delta R(f_R) + 8 \pi G \delta \rho_{\rm{M}} \right]
\end{equation}
as well as the counterpart of the Poisson equation 
\begin{equation}
	\label{eq:phi}
	\nabla^2 \Phi=\frac{16 \pi G}{3} a^2 \delta \rho_{\rm{M}} + \frac{a^2}{6} \delta R(f_R)
\end{equation}
can be easily obtained \citep{Li:2012bl,Zhao:2011bd}. Here $\delta R = R-\bar{R}$, $\delta \rho_{\rm{M}}=\rho_{\rm{M}}-\bar{\rho}_{\rm{M}}$ and $\Phi$ denotes the gravitational potential. The above equations are obtained in the quasi-static limit, neglecting time derivatives of the scalar field perturbation compared with the spatial derivatives. Equations \ref{eq:fr} and \ref{eq:phi} are closed given a functional form of $f(R)$, the density field and knowledge of the background evolution and represent the equations solved during our N-body simulations.

\subsection{Chameleon Mechanism}
Noting that $\nabla^2 \Phi_{\rm{GR}}=4 \pi G a^2 \delta \rho_{\rm{M}}$ and that in regions with low matter density we usually have $\delta R(f_R)\approx0$ we see that the equations thus decouple and become
ß
\begin{equation}
	\nabla^2 f_R = -\frac{2}{3} \nabla^2 \Phi_{\rm{GR}}
\end{equation}
\begin{equation}
	\nabla^2 \Phi=\frac{4}{3}\nabla^2 \Phi_{\rm{GR}}
\end{equation}
thus, in the underdense regime gravity is enhanced by a factor of $\frac{1}{3}$ relative to GR.

To pass local gravity tests, $f(R)$ gravity must be formulated such that GR is restored in high-density environments, such as the solar system, while leaving open the possibility of producing modified forces on large scales, which have lower matter density. Due in part to this transition between the screened and unscreened regimes, the model is in general inherently highly non-linear. In addition, as the modification is specifically chosen to match all solar system constraints, it should be very carefully compared with $\Lambda$CDM on these non-linear scales in order to distinguish between the two models. For example, \cite{Cardone:2012bd} constrain $f(R)$ models observationally using Type Ia Supernova and Gamma Ray Bursts, $H(z)$ data, BAO from SDSS \citep{2014ApJS..211...17A}, and WMAP7 data \citep{2011ApJS..192...18K}.

As the functional form of $f(R)$ is a free parameter we can conceivably pick this functional form to employ the ``chameleon mechanism'' in a high density environment. A Chameleon mechanism was first introduced cosmologically as a mechanism to give scalar fields an environment dependent effective mass \citep{Khoury:2003rn} allowing a scalar mediated force to be suppressed under certain environmental conditions. 

In this work we follow the Hu-Sawicki $f(R)$ chameleon model \citep{Hu:2007tv}:
\begin{equation}
	\label{eq:husawicki}
	f(R)=-m^2 \frac{c_1(-R/m^2)^n}{c_2(-R/m^2)^n+1}
\end{equation}
here $m^2\equiv 8 \pi G \bar{\rho}_{M,0}/3=H_0^2 \Omega_M$, $\Omega_M$ is the fractional matter density, $H_0$ the current Hubble expansion rate, and $n$, $c_1$ and $c_2$ are model parameters. 

In the Hu-Sawicki model we get $f_R \approx -\xi m^4 R^{-2}$. As $|R|\gg0$ becomes large, this function tends to zero. Thus the scalaron equation gives us $\delta R(f_R) = -8 \pi G \delta \rho_{\rm{M}}$. Plugging this into the Poisson equation, we see that we get 
\begin{equation}
	\nabla^2 \Phi = \nabla^2 \Phi_{\rm{GR}}
\end{equation}
Here we see in the dense regime where $f_R$ is close to zero we recover the Poisson equation for GR from \textbf{Equations \ref{eq:fr}} and \textbf{\ref{eq:phi}}. Thus GR is restored in dense regions, as desired. 

Examining the behavior in more detail, we see that observations require the absolute value of the scalaron today should be sufficiently small. In the Hu-Sawicki model, we can examine how to ensure it matches the $\Lambda$CDM background evolution and fits with the present constraint on the value of the background field at $z=0$, $|f_{R0}|$ by fixing the free model parameters. In the background the scalaron always sits close to the minimum of the effective potential that governs its dynamics as the background cosmology implies the particle masses and the gravitational constant cannot vary substantially between Big Bang Nucleosynthesis (BBN) and now \citep{Brax:2012tq}.

This effective potential can be derived by taking the trace of the modified Einstein equation in $f(R)$ theory:
\begin{equation}
	\square f_R = \frac{\partial V_{eff}}{\partial f_R}= \frac{1}{3}(R-f_R R+2f+8 \pi G \rho_M)
\end{equation}
setting this value to 0 and as $f_R \approx 0$ in the background, we obtain:
\begin{equation}
	%TODO: this is opposite sign of R to Zhao!! se eq. 12
	\bar{R}\approx 8\pi G \bar{\rho}_M - 2 \bar{f} \approx 3m^2(a^{-3}+\frac{2}{3}\frac{c_1}{c_2})
\end{equation}
tuning this to match the $\Lambda$CDM background evolution we obtain $\frac{c_1}{c_2}=6 \frac{\Omega_\Lambda}{\Omega_m}$. Plugging in values from WMAP5 for $\Omega_\Lambda$ and $\Omega_m$,  0.728 and 0.272  respectively, we see that $\bar{R}\simeq 41 m^2 \gg m^2$ meaning we can simplify \textbf{Equation \ref{eq:husawicki}} to read
\begin{equation}
	\label{eq:frsimple}
	f_R \approx -\frac{n c_1}{c_2^2}(\frac{m^2}{-R})^{n+1}
\end{equation}
We see from Equation \ref{eq:frsimple} that we can bundle $c_1$ and $c_2$ into a single free parameter $\xi = \frac{c_1}{c_2^2}$. Deriving the relation to the scalaron today we obtain
\begin{equation}
	\xi= -\frac{1}{n}\left[ 3(1+4\frac{\Omega_\Lambda}{\Omega_m})\right]^{n+1} f_{R0}
\end{equation}
and we can see that this parameter can be derived uniquely from $\xi$ and has a ready physical interpretation. Thus $c_1/c_2$ determines the expansion rate of the universe and $\xi$ determines the structure formation.
%\cite{Capozziello:2007bf} constrain $n>0.9$ by prohibiting violations of the strong and weak equivalence principles

In practice $n$ is often taken to be an integer. In this paper we concentrate on models with $n=1$ for convenience. We consider three representative choices of the other Hu-Sawicki model parameter, with $|f_{R0}|=10^{-4}$, $10^{-5}$ and $10^{-6}$, which we call model F4, F5 and F6 respectively. The relation between $|f_{R0}|$ and the corresponding Compton wavelength of the scalaron field, $\lambda_C$, is approximately $\lambda_C=32\sqrt{|f_{R0}|/10^{-4}}$ Mpc. The Compton wavelength corresponds roughly to the range of the fifth force, beyond which the it decays quickly. We find that at the present day $\lambda_C$ is about 32, 10 and 3 Mpc for F4, F5 and F6 respectively. Therefore, F4 has the strongest deviation from GR while F6 has the weakest fifth force.

\subsection{$f(R)$ gravity in simulation}
There are two primary ways to constrain modified gravity models at the cosmological scale by observations. Firstly are methods which test the effects on the cosmic expansion history such as measurements of baryonic acoustic oscillations (BAO) or Type Ia supernovae (SNIa) most saliently. These constraints are difficult, as in general the background expansion history of the $f(R)$ models studied here is expected to be nearly identical with that of $\Lambda$CDM. Secondly are constraints from measurements of the cosmic growth history, through observations of e.g. weak lensing, galaxy cluster behavior, or the Integrated Sachs Wolfe (ISW) effect \citep{Lombriser:2011hi,Mak:2012io,Lombriser:2013wta,Dossett:2014cy,Zhang:2012vm}. Authors have recently deployed numerical simulations in the linear or ``no-chameleon'' regimes \citep{Lombriser:2011hi,Lombriser:2012cf} to develop constraints in tandem with such observations. Constraints relying only indirectly on simulation are possible, for example those which use a parameterized Post-Friedmann (PPF) framework and linear theory \citep{Lombriser:2011hi,Lombriser:2013wta} or deploy analytic results motivated by simulation \citep{Terukina:2014jq}. 

These constraints are  promising, but more accurate theoretical predictions require solving the full non-linear equations in simulation. Thus, methods to explore the non-linear regime in $f(R)$ gravity are of high theoretical importance. Particularly the non-linear scales are critical for weak lensing measurement signals, and lend themselves to detailed observational comparisons. Moreover, quantifying the chameleon effect in details enables discrimination between different $f(R)$ models themselves. Yet, to date due to the difficulty solving the coupled scalar field and modified Poisson equations, it has not been straightforward to explore at high resolution these consequences and to adequately be assured of convergence and good statistics.

In recent years there has been work to modify existing N-body codes to support $f(R)$ gravity and similar theories. Due to the highly non-linear nature of the equations, early efforts have largely been limited in resolution and scale \citep{Oyaizu:2008gb,Schmidt:2009cn,Schmidt:2009iy,Li:2009ix,Brax:2012jv,Ferraro:2011gn}.
Recent codes have been able to explore the non-linear effects of $f(R)$ gravity on an unprecedented level \citep{Li:2012bl,2013MNRAS.436..348P}. In this paper we work with a modification to the adaptive mesh refinement (AMR) cosmological simulation code \texttt{RAMSES} \citep{Teyssier:2002fj}, \texttt{ECOSMOG} introduced by \cite{Li:2012bl}, integrating the ability to solve the equations for the $f(R)$ scalar field on an AMR grid and enabling higher resolution simulations than previously possible. Several research programs have thus far been carried out using \texttt{ECOSMOG}. For example, \cite{Jennings:2012tu} study the clustering of dark matter in redshift space in $f(R)$ gravity models, finding a significant deviation from $\Lambda$CDM. \cite{Hellwing:2013rxa}  study high-order clustering, \cite{Zu:2013joa} study galaxy infall kinematics, and \cite{2013arXiv1310.6986C} study the ISW effect using \texttt{ECOSMOG} in the context of $f(R)$ gravity respectively.

However, these previous studies are still limited by resolution effects. In \cite{Li:2012ck} it is shown that $f(R)$ simulations with larger boxes and lower resolutions systematically underestimate the density field, over estimate the contribution of the fifth force, and overestimate power on small scales--predicting a greater clustering of matter than a higher resolution simulation. This is due to the fact that the fifth force becomes weak in exactly the high density regions in which high resolution is required. Moreover, it is well known that lower resolution simulations can be subject to the over-merging problem particularly in dense environments such as the cluster environment of our work \citep{1999ApJ...516..530K}. 
%This problem disappears at higher resolution in $\Lambda$CDM simulations, so can be expected to behave similarly for the F6 model, which closely follows the evolution of $\Lambda$CDM, and to some extent for the F5 and F4 models as well.

%, 2012MNRAS.422.3081M,ramses
%,2013MNRAS.436..460W,2014ApJS..210...14K, other community codes
To address resolution effects, we are able to ensure our simulations cover a wide range of length and mass scales, and thus extend and refine results obtained at lower resolution. For example, \cite{2013ApJ...779...39J} develop constraints on $f(R)$ gravity using a sample of unscreened dwarf galaxies deploying criteria based on lower resolution simulations by \cite{Zhao:2011bd} and \cite{Cabre:2012cm}, applying a screening criterion motivated by numerical simulation results to observational data. We explore the validity of the \cite{Cabre:2012cm} criteria in the context of the satellites which we are able to resolve in our high resolution suite of simulations.

To achieve this goal and mitigate these resolution effects allowing us to focus on cluster scale properties, we deploy a zoom technique, running a lower resolution simulation and re-simulating an area of interest at higher resolution. This technique has been successfully deployed in the past decades to focus resolution where it is desired \citep{1997MNRAS.286..865T,1999ApJ...524L..19M,2005MNRAS.359.1537R,2005MNRAS.363..379G}. 

In recent years, following a similar research technique has been highly successful in the \texttt{RAMSES} community \citep{2004MNRAS.349.1039N,2010MNRAS.405..274H,2012MNRAS.420.2859M,Martizzi:2013aja,2013arXiv1308.6321R,Martizzi:2014ka} and in other community codes \citep{2011MNRAS.415.2101H,2013ApJ...763...70W,2014MNRAS.437.1894O,2014arXiv1407.7129D}. In this work we explore for the first time \texttt{ECOSMOG} in a zoom simulation mode. Simulating our cluster sized area of physical interest in a cosmological context is enabled through a multi-scale series of initial conditions with AMR being run only at the inner most level. Using this technique in the \texttt{ECOSMOG} context, we are able to zoom in on the properties of a halo of interest in high resolution while maintaining the adequate cosmological boundary conditions, and present the detailed evolution of a Virgo like $10^{14} M_\odot$ mass halo under $f(R)$ gravity simulated to unprecedented resolution.

\section{Simulations}
\label{sec:simulation}
For our simulations we provided initial conditions computed using the Eisenstein \& Hu transfer function \citep{Eisenstein:1998ez} computed using the \texttt{Grafic++} code \citep{Potter:bWn-q5m9} as input to the \texttt{RAMSES} code \citep{Teyssier:2002fj}. We performed a suite of dark matter only zoom cosmological simulations, each using the common cosmological parameters set using WMAP5 results as listed in \textbf{Table \ref{cosmologyparamstable}}, in a set of $f(R)$ models with varying parameters listed in \textbf{Table \ref{simulationdesc}}. The zoom technique selects a subregion of the computational domain to focus on to achieve the desired resolution. For this work we chose to focus on a halo of cluster size mass.

\begin{table*}
 \centering
 \begin{minipage}{140mm}

\caption{Cosmological parameters for our simulations.}
\label{cosmologyparamstable}
  \begin{tabular}{ c | c | c| c | c | c | c | c }
    \hline
     \textit{$H_0$} [km $\rm{s}^{-1} \rm{Mpc}^{-1}$] & \textit{$\sigma_8$} & \textit{$n_s$} & \textit{$\Omega_\Lambda$} & \textit{$\Omega_m$} & \textit{$\Omega_b$} & $\mathit{m_{cdm}} [10^6 M_\odot]$ & $\mathit{\Delta x_{\rm{min}}} \rm[kpc/h]$\\
    \hline
    70.4 &  0.809 & 0.809 & 0.728 & 0.272 & - & 36 & 8.04  \\
    \hline
  \end{tabular}
  \end{minipage}
\end{table*}

\begin{table*}
 \centering
 \begin{minipage}{140mm}
\caption{Hu-Sawiki $f(R)$ model and simulation parameters}
\label{simulationdesc}
 \begin{tabular}{ c | c | c| c | c | c}
    \hline
    Simulation & $n$ &  $\frac{c_1}{c_2^2}$ & pre-smoothing (fine/course)  & post-smoothing (fine/course) & $\lambda_C$ [Mpc] \\
    \hline
    $\Lambda$CDM & - & - & - & - & - \\
    $|f_{R0}|=10^{-6}$ (F4) & 1 & 0.168 & 1000/100 & 10/10  & 3\\ 
    $|f_{R0}|=10^{-5} $(F5) & 1 & 0.0168 & 1000/100 & 10/10 & 10 \\
    $|f_{R0}|=10^{-4}$ (F6) & 1 & 0.00168 & 1000/100 & 10/10 & 32 \\
    \hline
  \end{tabular}
  \end{minipage}
  \end{table*}

To use the zoom simulation technique, a simulation was first run in uniform resolution (unigrid mode) at a resolution of $128^3$, where we determined the region of interest, our Virgo like halo. To select the zoom subregion, in this lower resolution simulation we identified dark matter halos and subhalos using the \texttt{AdaptaHOP} algorithm \citep{Aubert:2004im} and a merger tree identification algorithm, implemented in the \texttt{GalICs} pipeline \citep{Tweed:2009fh}. Using the catalog of structures identified, we selected candidate halos of mass $\simeq 10^{14} M_\odot$. Our best candidate would have a quiescent merger history and a stable NFW profile at $z=0$ in the $\Lambda$CDM simulation. We used the merger trees to examine the assembly history to select exactly such a halo. The virial radius of our selected halo, which we define as $r_{200 \rho_c}$ is $\simeq 1$Mpc$h^{-1}$ ; the virial mass of our selected halo, which we define as $\rm{M}_{200 \rho_c}$, is $\simeq 1\times 10^{14} M_\odot$ and its last major merger occurs at $z\sim 1.5$. 

Once selected, a quasi-spherical region was determined which encapsulated all the particles ending up in the final selected halo by taking the Lagrangian volume of the particles we are interested at $z=0$ in the initial conditions, and further more adding in additional particles in the boundary region. Then we generated a new set of initial conditions using \texttt{Grafic++} providing the same large scale modes but at higher resolution. Generating a nested set of rectangular grids around this region with various particle masses, in our case to reach our desired resolution at a zoom of 50 Mpc$h^{-1}$ in a 200Mpc$h^{-1}$ initial box, we had a series of 5 nested grids to reach a final effective resolution of $2048^3$. 

Going from a coarse level to a finer level, it was always ensured that at least 10 boundary cells stood between each level, and AMR was only run in the highest resolution region. Here the max level of refinement in the zoom AMR simulation was 17-18 depending on the model. For the solvers, a standard Dirichlet boundary condition was used. A test as to accuracy is contamination of the final region by higher mass particles; we saw in this region zero contamination.

Four sets of high resolution zoom simulations were run, for the $\Lambda$CDM, F4, F5, and F6 models respectively. Their model parameter details are given in \textbf{Table \ref{simulationdesc}} and their cosmological parameters are given in \textbf{Table \ref{cosmologyparamstable}}.

\section{Overview}
\label{sec:visual}
We computed standard surface density distribution and line of sight velocity dispersion profile for our simulations. Each of these quantities has a dependence on chosen line of sight for the measurement. To quantify this dependency, for all of our analysis we randomized over the lines of sight and error bars indicate the scatter in the measured quantity over 100 such random lines of sight and the results are detailed in this section.

\begin{figure*}
\begin{center}$
\begin{array}{ccccc}
 z & F4 & F5 & F6 & \Lambda CDM \\
3.9 &
\includegraphics[keepaspectratio,width=0.16\textwidth]{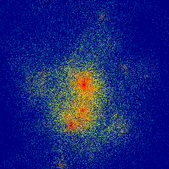} &
\includegraphics[keepaspectratio,width=0.16\textwidth]{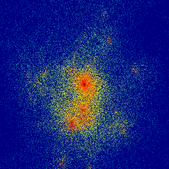} &
\includegraphics[keepaspectratio,width=0.16\textwidth]{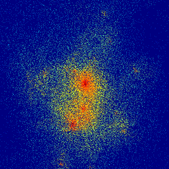} &
\includegraphics[keepaspectratio,width=0.16\textwidth]{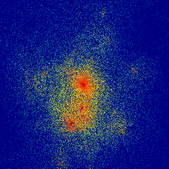} \\

 2.3 &
\includegraphics[keepaspectratio,width=0.16\textwidth]{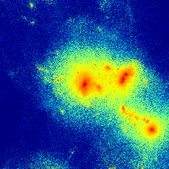} &
\includegraphics[keepaspectratio,width=0.16\textwidth]{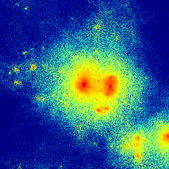} &
\includegraphics[keepaspectratio,width=0.16\textwidth]{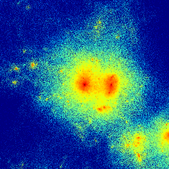} &
\includegraphics[keepaspectratio,width=0.16\textwidth]{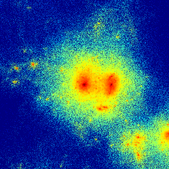} \\

1.5 &
\includegraphics[keepaspectratio,width=0.16\textwidth]{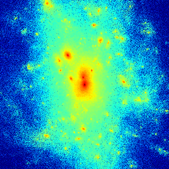} &
\includegraphics[keepaspectratio,width=0.16\textwidth]{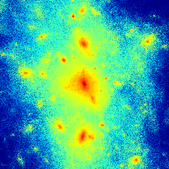} &
\includegraphics[keepaspectratio,width=0.16\textwidth]{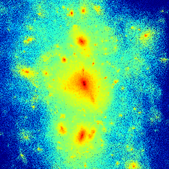} &
\includegraphics[keepaspectratio,width=0.16\textwidth]{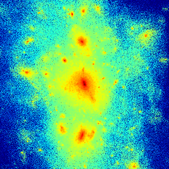} \\

1 &
\includegraphics[keepaspectratio,width=0.16\textwidth]{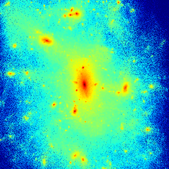} &
\includegraphics[keepaspectratio,width=0.16\textwidth]{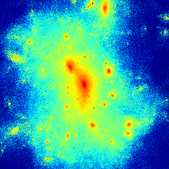} &
\includegraphics[keepaspectratio,width=0.16\textwidth]{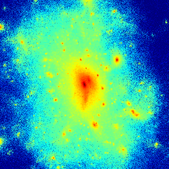} &
\includegraphics[keepaspectratio,width=0.16\textwidth]{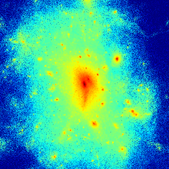} \\

0.66 &
\includegraphics[keepaspectratio,width=0.16\textwidth]{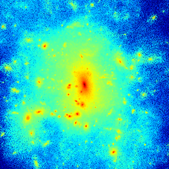} &
\includegraphics[keepaspectratio,width=0.16\textwidth]{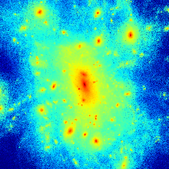} &
\includegraphics[keepaspectratio,width=0.16\textwidth]{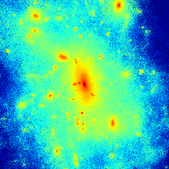} &
\includegraphics[keepaspectratio,width=0.16\textwidth]{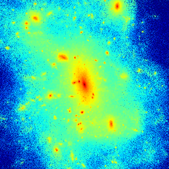} \\

0.4 &
\includegraphics[keepaspectratio,width=0.16\textwidth]{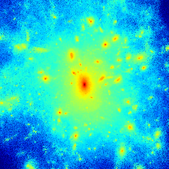} &
\includegraphics[keepaspectratio,width=0.16\textwidth]{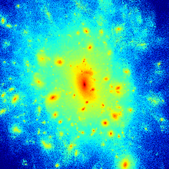} &
\includegraphics[keepaspectratio,width=0.16\textwidth]{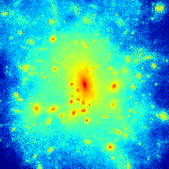} &
\includegraphics[keepaspectratio,width=0.16\textwidth]{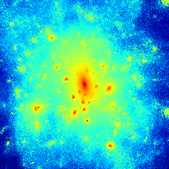} \\

0.25 &
\includegraphics[keepaspectratio,width=0.16\textwidth]{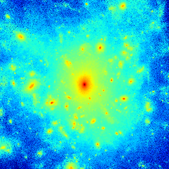} &
\includegraphics[keepaspectratio,width=0.16\textwidth]{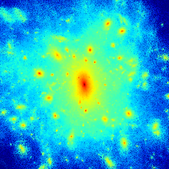} &
\includegraphics[keepaspectratio,width=0.16\textwidth]{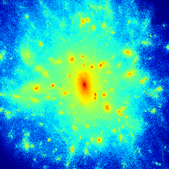} &
\includegraphics[keepaspectratio,width=0.16\textwidth]{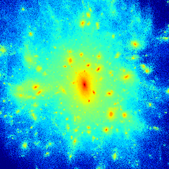} \\

0.1 &
\includegraphics[keepaspectratio,width=0.16\textwidth]{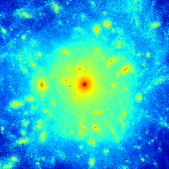} &
\includegraphics[keepaspectratio,width=0.16\textwidth]{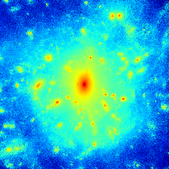} &
\includegraphics[keepaspectratio,width=0.16\textwidth]{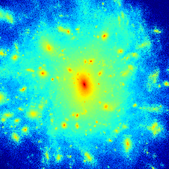} &
\includegraphics[keepaspectratio,width=0.16\textwidth]{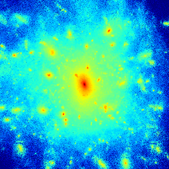} \\

0 &
\includegraphics[keepaspectratio,width=0.16\textwidth]{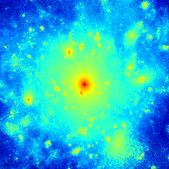} &
\includegraphics[keepaspectratio,width=0.16\textwidth]{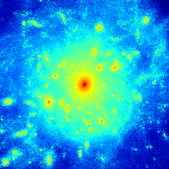} &
\includegraphics[keepaspectratio,width=0.16\textwidth]{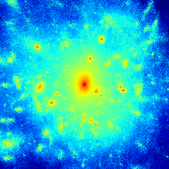} &
\includegraphics[keepaspectratio,width=0.16\textwidth]{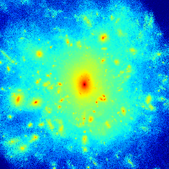} \\

\end{array}$
\end{center}
\caption{Density along the line of sight is depicted for a visual comparison of the four models at various redshifts.}
\label{visual}
\end{figure*}

As a result of the highly non-linear nature of the coupled Poisson and scalar field equations, the strength of the chameleon effect  (i.e., the screening of the fifth force) and the regions where gravity is enhanced depend on both the radius from the main halo center and the redshift. To get a sense of these effects, we explore a visual comparison of the four models at various redshifts. This is depicted in \textbf{Figure \ref{visual}}. Here we can see the F4 model, the model in which the modification is strongest at $z=0$, already at high redshift has increased structure formation, compared to other models. By $z=0.4$ the main halo in the F4 model has accreted most of its surrounding subhalos, while for the weaker models this accretion is ongoing at lower $z$. We will explore this behavior in the context of the strength of the force modifications.

\section{Profiles across Halos, Models, and Resolution}
\label{sec:profiles}

\begin{figure}
\centering
\caption{Cylindrical surface density profile. Error bars come from the variation of the quantity of 100 random lines of sight}
\label{sdraw}
\begin{subfigure}{0.95\linewidth}
  \centering \includegraphics[keepaspectratio,width=0.95\linewidth]{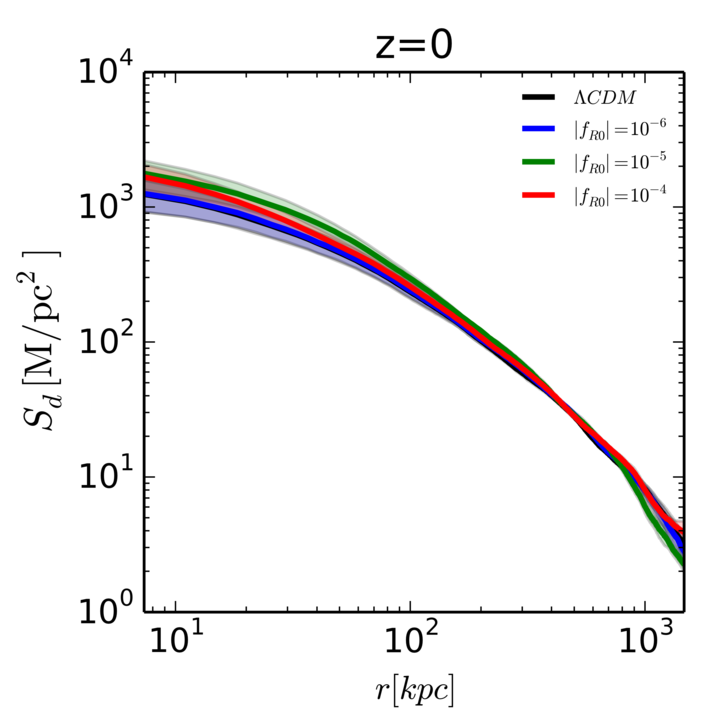}
\end{subfigure}
\begin{subfigure}{0.95\linewidth}
  \centering  \includegraphics[keepaspectratio,width=0.95\linewidth]{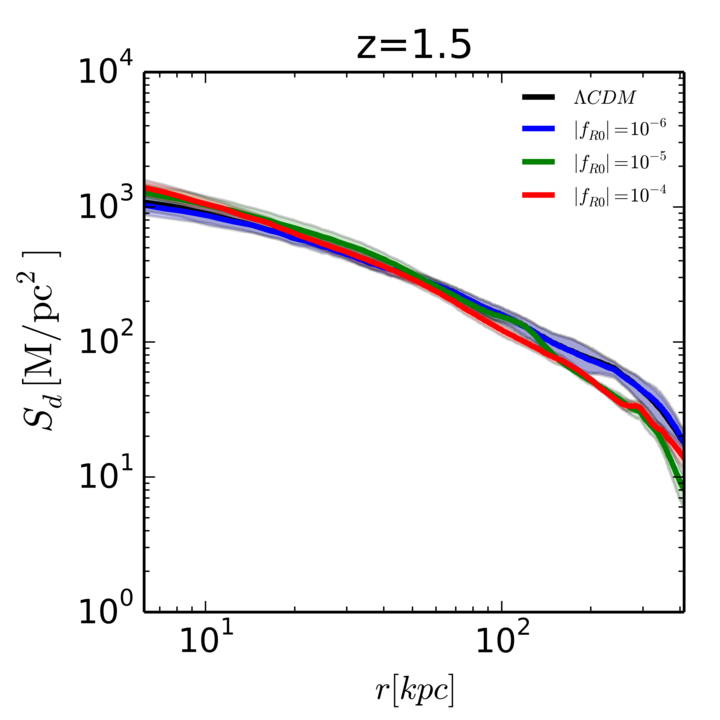}
\end{subfigure}
\label{sdsub}
\end{figure}

\begin{figure}
\centering
\caption{Cylindrical velocity dispersion profile. Error bars come from the variation of the quantity of 100 random lines of sight}
\label{szraw}
\begin{subfigure}{0.95\linewidth}
  \centering \includegraphics[keepaspectratio,width=0.95\linewidth]{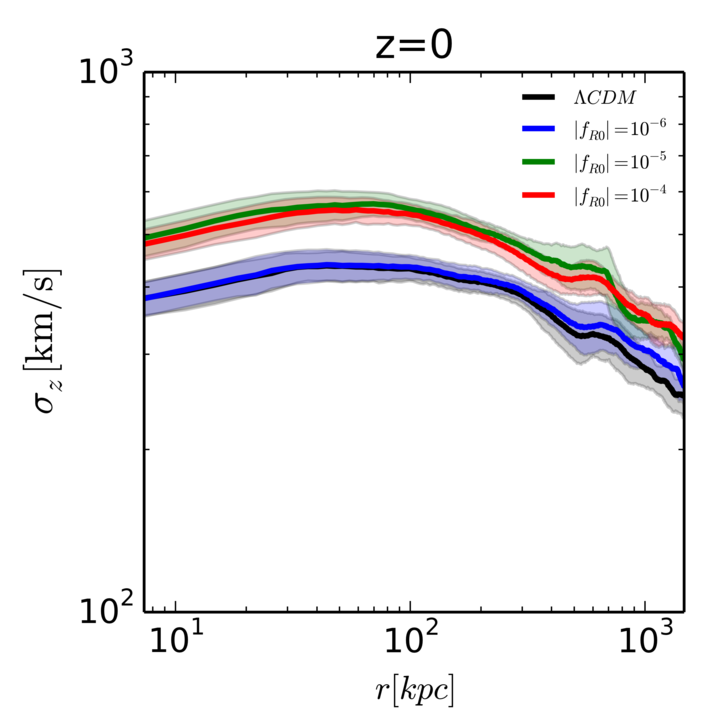}
\end{subfigure}
\begin{subfigure}{0.95\linewidth}
  \centering \includegraphics[keepaspectratio,width=0.95\linewidth]{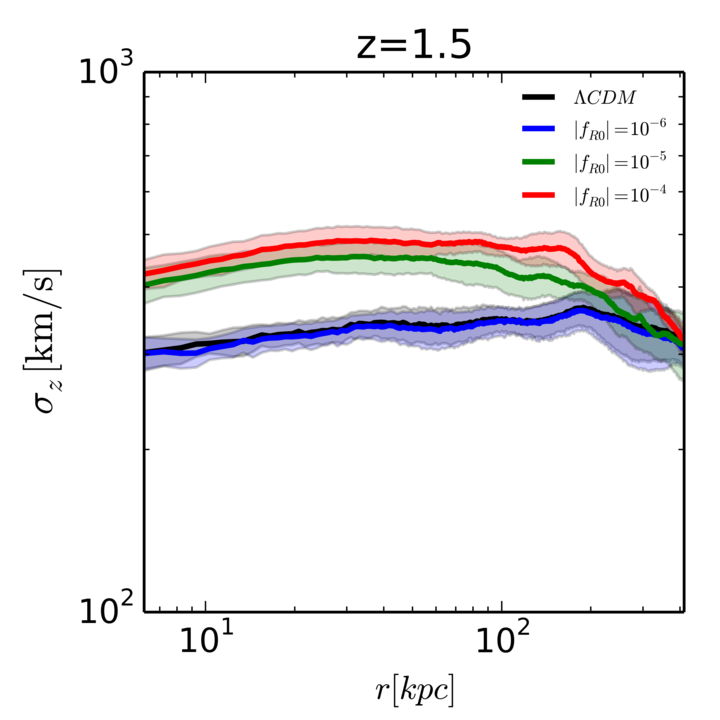}
\end{subfigure}
\label{szsub}
\end{figure}

As shown in \cite{Lombriser:2013wta}, the $f(R)$ halos are well fit by an NFW profile as in $\Lambda$CDM. To see the raw differences the cylindrical surface density profile and velocity dispersion profiles are presented at $z=0$ and $z=1.5$ in \textbf{Figure \ref{sdsub}} and \textbf{Figure \ref{szsub}} respectively. Of the two, the velocity dispersion profile is where we see the maximum effect of the $|f_{R0}|=10^{-6}$ model. Here error bars come from the variation of that quantity over our 100 random lines of sight. 

We see clear differences in the profiles at both redshifts, with the stronger $f(R)$ models in general having greater velocity dispersion and a higher central density, or concentration, than the weaker models. We can already see in this limited sample of the simulation that the relative behavior of the various models have a complicated dependence on model strength. Considering the velocity dispersion profile for example, at $z=0$ there is a clear discrimination between the pair of weak/no $f(R)$ models and the pair of strong $f(R)$ models at all radii, while at higher redshift these models become completely degenerate at higher radius. Considering the surface density profile, for example at $z=1.5$ the strongest $f(R)$ model F4 has the highest concentration while at $z=0$ the weaker F5 model has the highest concentration. 

We focus on the cluster profile, while \cite{Li:2012ck} focus on the power spectra, as in \cite{Li:2012ck} we find that the velocity (peculiar velocity field power spectrum was rather analyzed in \cite{Li:2012ck}) is more affected by the presence of a fifth force than the surface density distribution (matter power spectrum was rather analyzed in their case). Likewise, as in \cite{Li:2012ck} we find different effects of modified gravity, in general, on the velocity dispersion profile as compared to the surface density distribution profile. Finally, as in \cite{Li:2012ck} we find that across models the shape and evolution although depending on $f(R)$ model and time follow roughly the same sequence, but for models with smaller $|f_{R0}|$ the evolution is delayed due to the suppression of the fifth force until comparatively later times.

Viewing the profiles gives us a picture of the differences between the various models and their associated degeneracies, but quantitative differences are more apparent in the relative difference plots by comparing the ratio of the values of a modified gravity model to that of $\Lambda$CDM. In identical models this would be 1, so by comparing this value to 1 we can in effect see the strength of the modification of the profile as compared to $\Lambda$CDM. 

\begin{figure*}
\begin{center}$
\begin{array}{ccccc}
\includegraphics[keepaspectratio,width=0.3\linewidth]{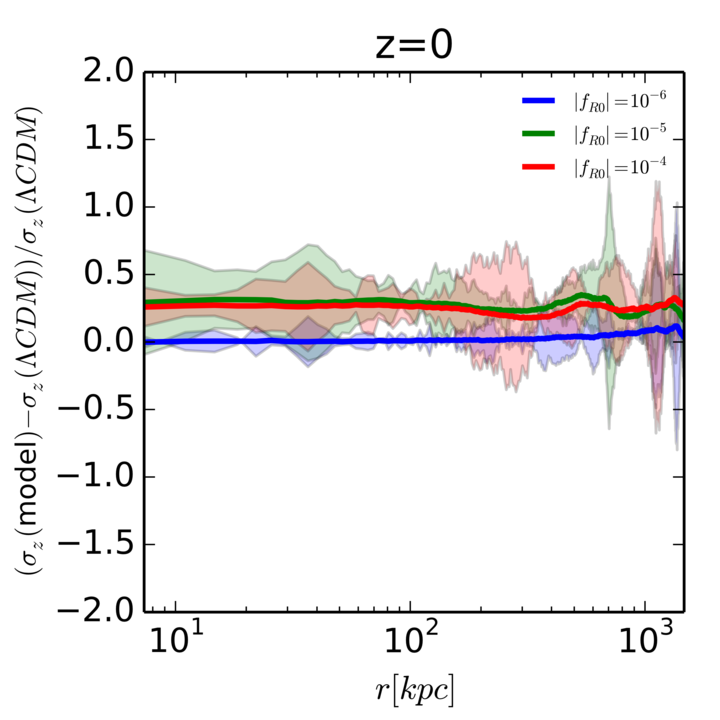} &
\includegraphics[keepaspectratio,width=0.3\linewidth]{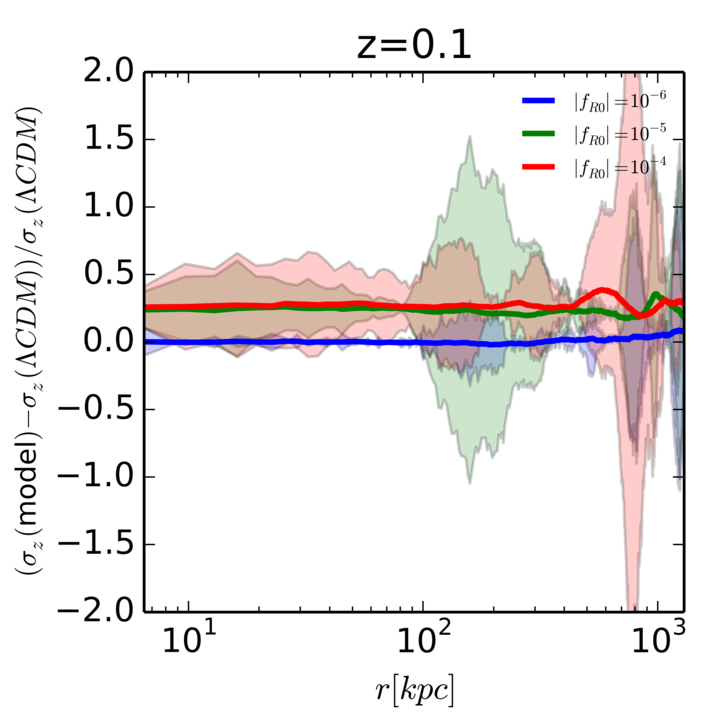} &
\includegraphics[keepaspectratio,width=0.3\linewidth]{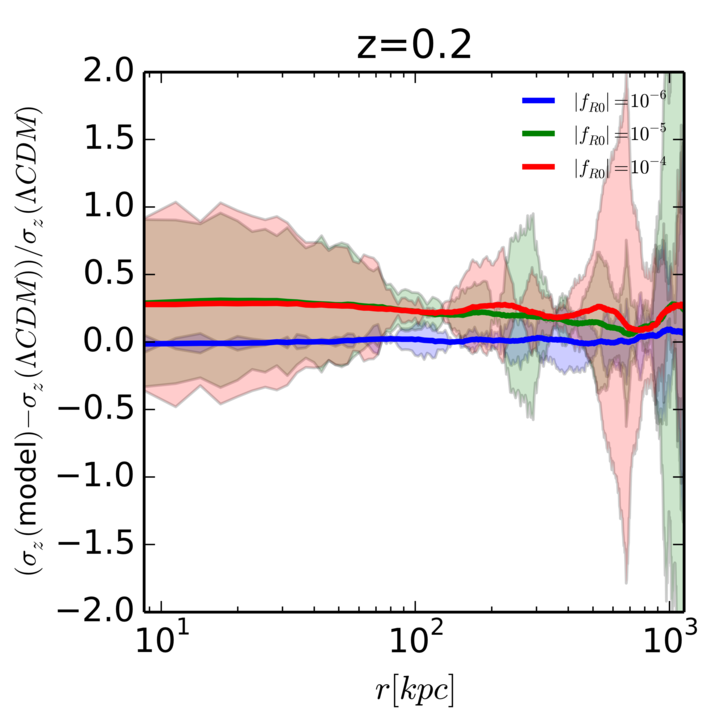} \\
\includegraphics[keepaspectratio,width=0.3\linewidth]{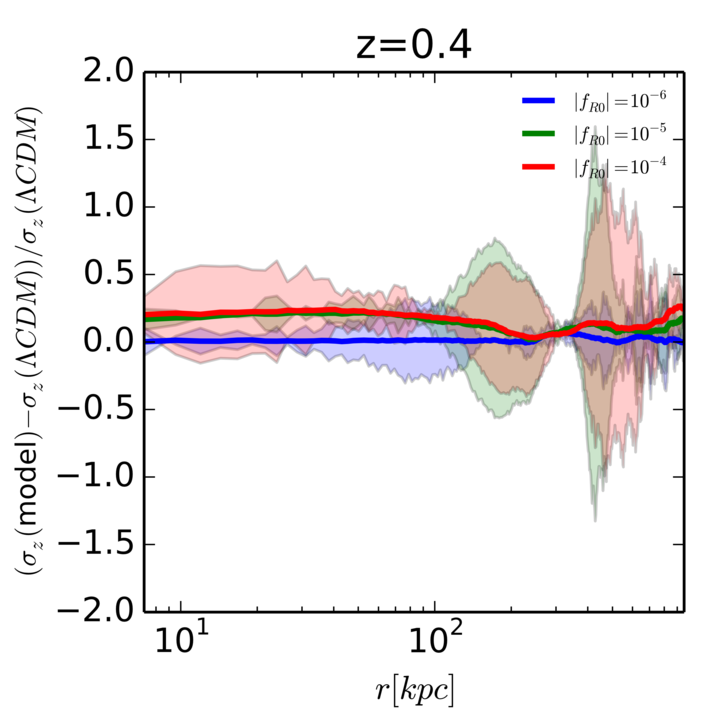} &
\includegraphics[keepaspectratio,width=0.3\linewidth]{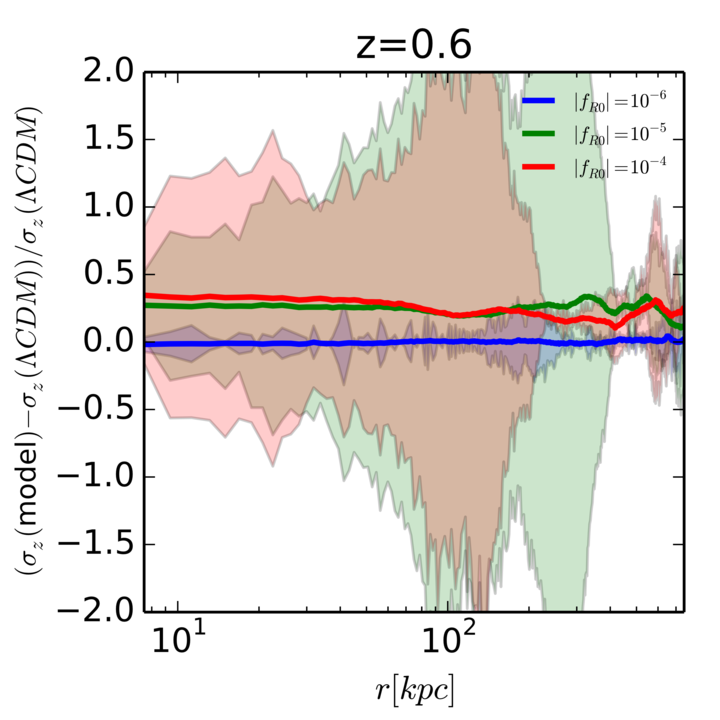} &
\includegraphics[keepaspectratio,width=0.3\linewidth]{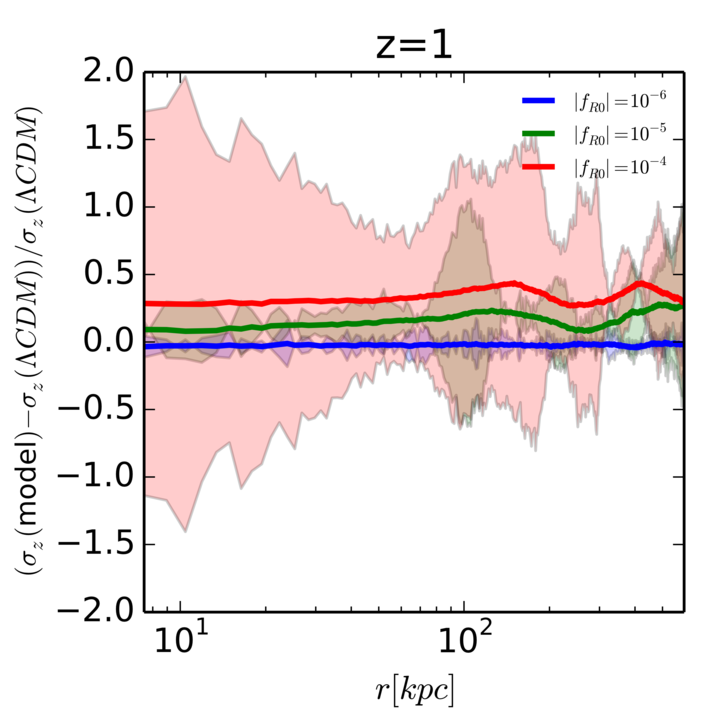} \\
\includegraphics[keepaspectratio,width=0.3\linewidth]{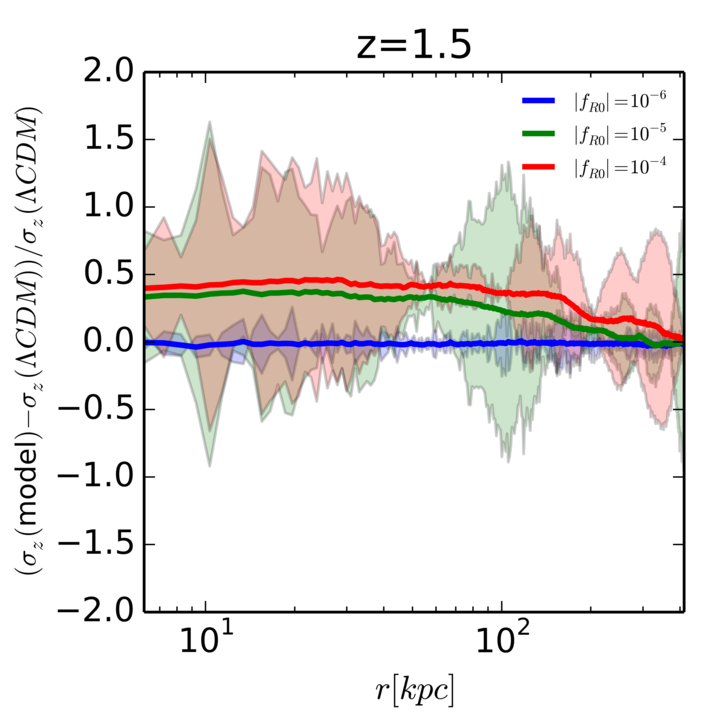} &
\includegraphics[keepaspectratio,width=0.3\linewidth]{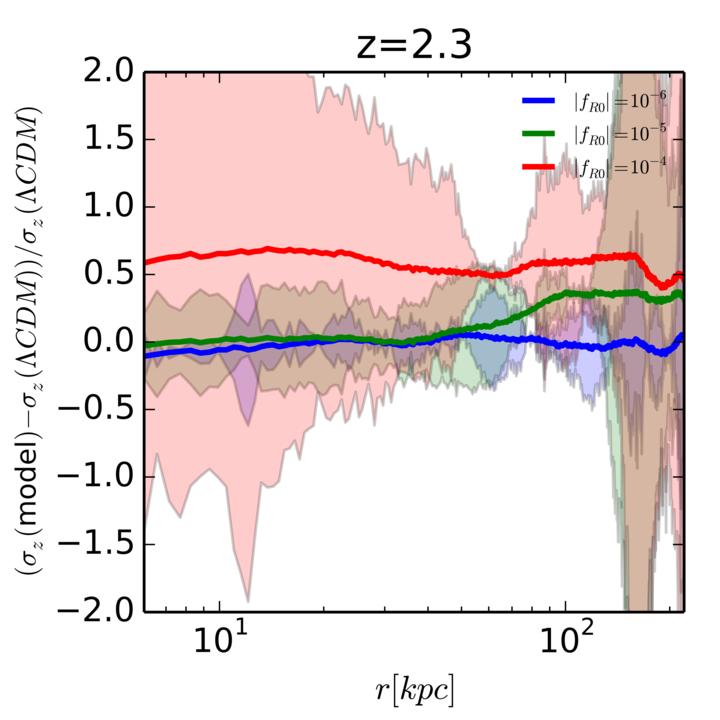} &
\includegraphics[keepaspectratio,width=0.3\linewidth]{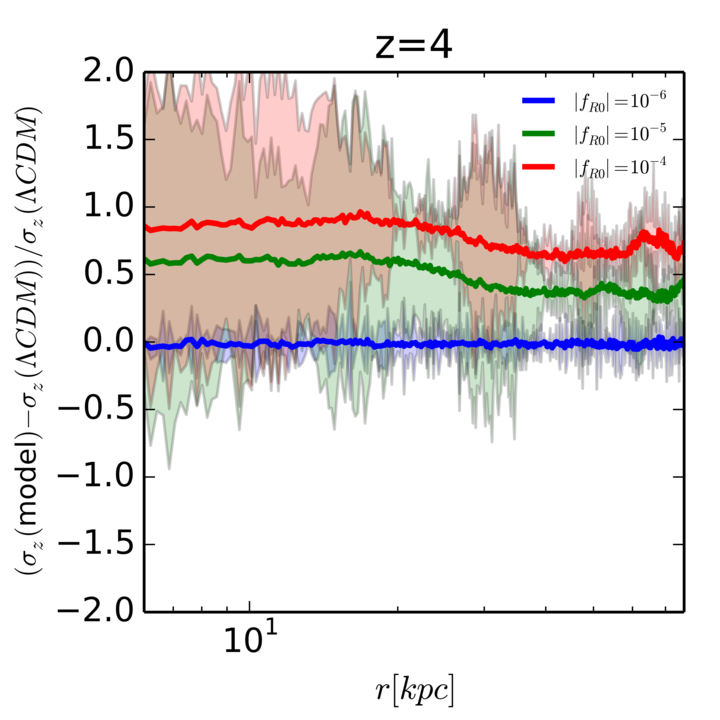} 
\end{array}$
\end{center}
\caption{Relative differences of cylindrical velocity dispersion profile to fiducial model. Error bars come from the variation of the quantity of 100 random lines of sight.}
\label{sigmazdiff}
\end{figure*}

The relative velocity dispersion profile is depicted in \textbf{Figure \ref{sigmazdiff}} where we can see in general the weakest F6 model is highly degenerate with $\Lambda$CDM, having only a small deviation in the error bars over chosen line of sight at intermediate radius and redshift. In contrast, the F5 and F4 models are generally enhanced in velocity at all redshifts in a distinguishable manner from the $\Lambda$CDM and F6 models. Compared with each other, the F5 and F4 are degenerate at low redshift, but become distinguishable at higher redshift, greater than $z=1$ in our plots. Their distinguishability remains roughly constant with radius.

At late times, the velocity dispersion of F4 and F5 is consistently at about 4/3 times the GR value, which makes sense considering the fact that $G$ is rescaled by 4/3 in $f(R)$ gravity. In F6 the difference from GR is much smaller primarily as the particles have never experienced the enhanced gravitational strength. For particles that have not fallen into the halo at early times, the fifth force is weak because it is generally weak at high redshift, and after they have fallen into the halo the fifth force is weak again because of the screening. The fact that F4 and F5 are almost indistinguishable below $z=1$ is because the fifth force can at most be 1/3 of standard gravity, and for both models it has achieved this upper bound. Finally it is interesting to note that the enhancement of the velocity dispersion at low redshifts in F4 and F5 is almost independent of the distance from the halo centre.

\begin{figure*}
\begin{center}$
\begin{array}{ccccc}
\includegraphics[keepaspectratio,width=0.3\linewidth]{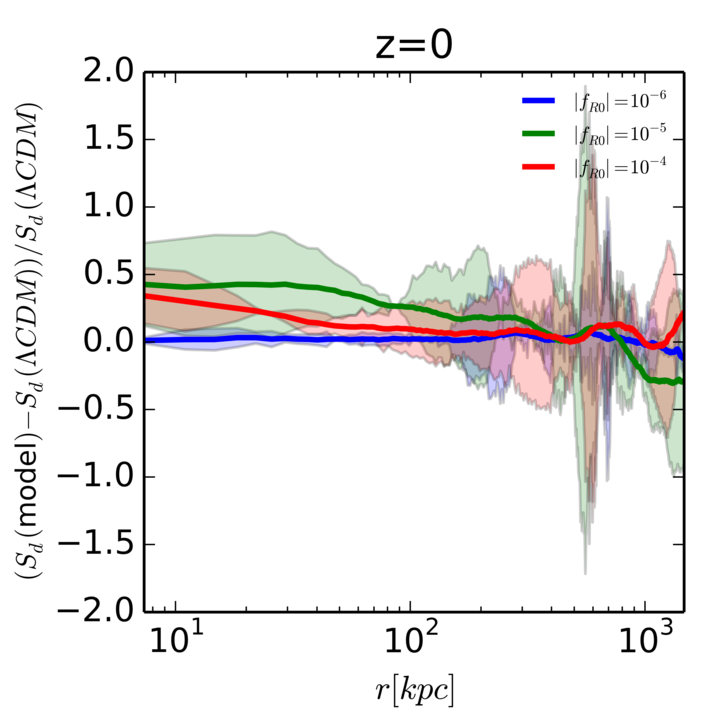} &
\includegraphics[keepaspectratio,width=0.3\linewidth]{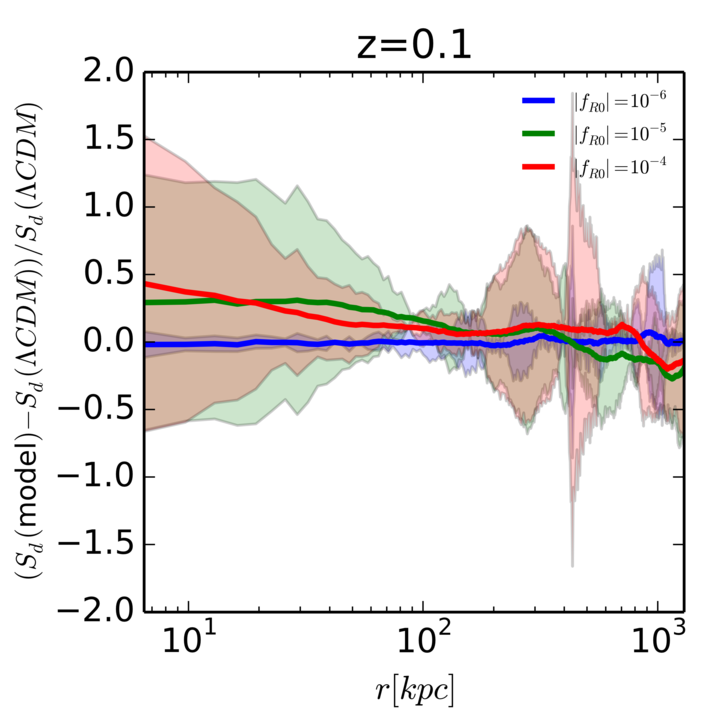} &
\includegraphics[keepaspectratio,width=0.3\linewidth]{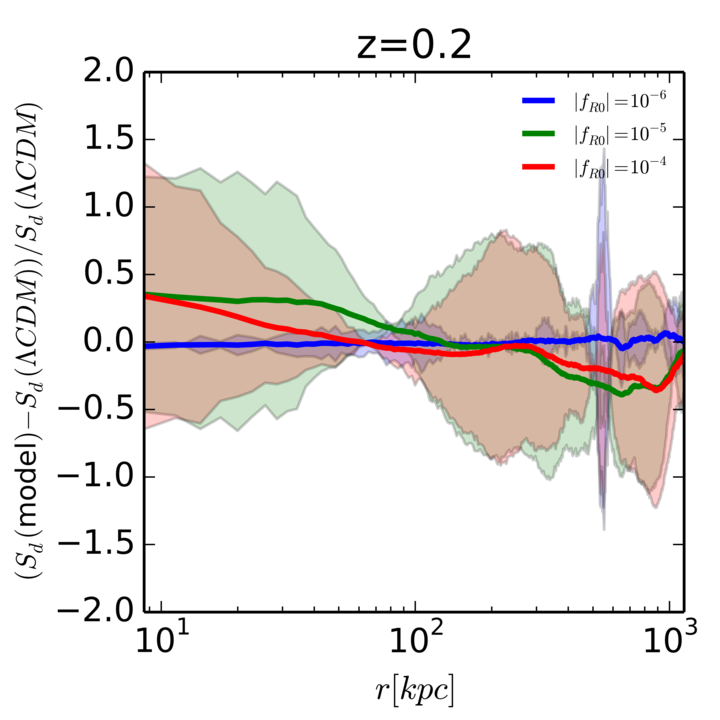} \\
\includegraphics[keepaspectratio,width=0.3\linewidth]{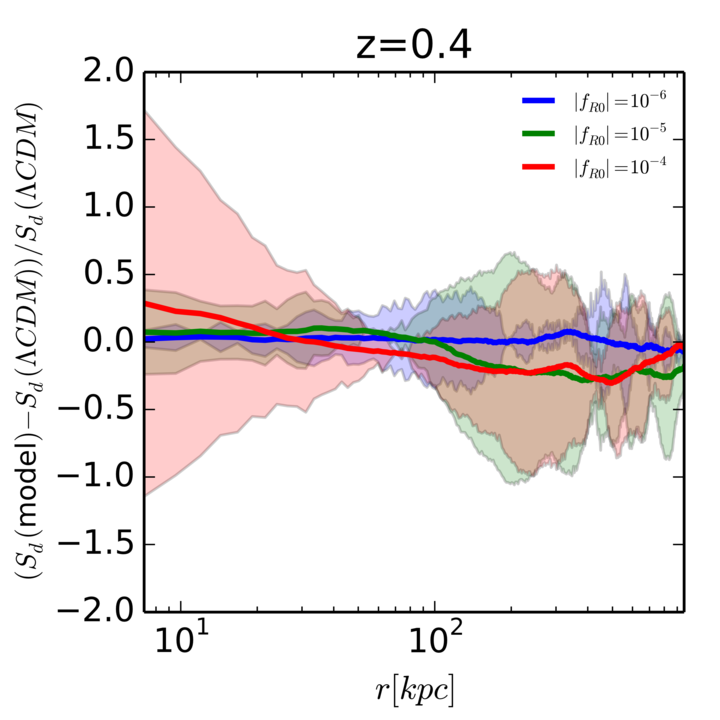} &
\includegraphics[keepaspectratio,width=0.3\linewidth]{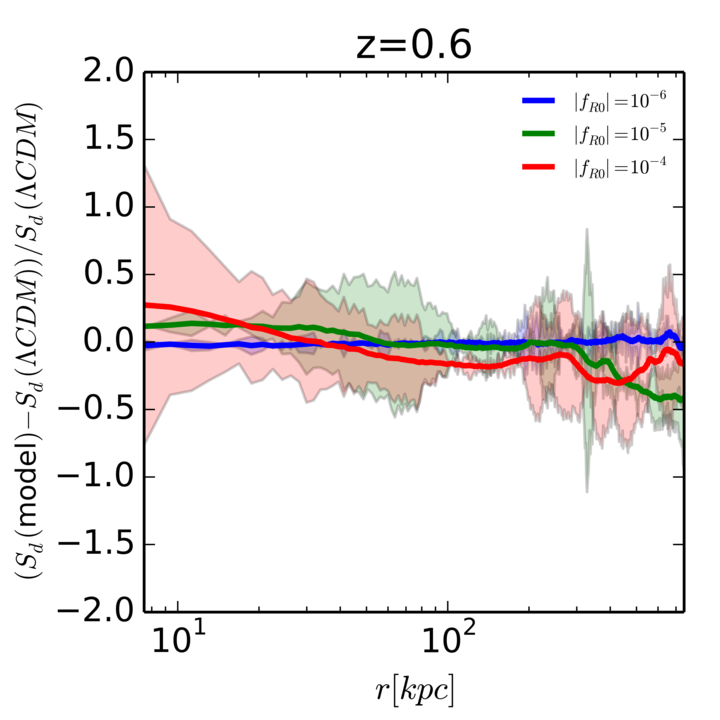} &
\includegraphics[keepaspectratio,width=0.3\linewidth]{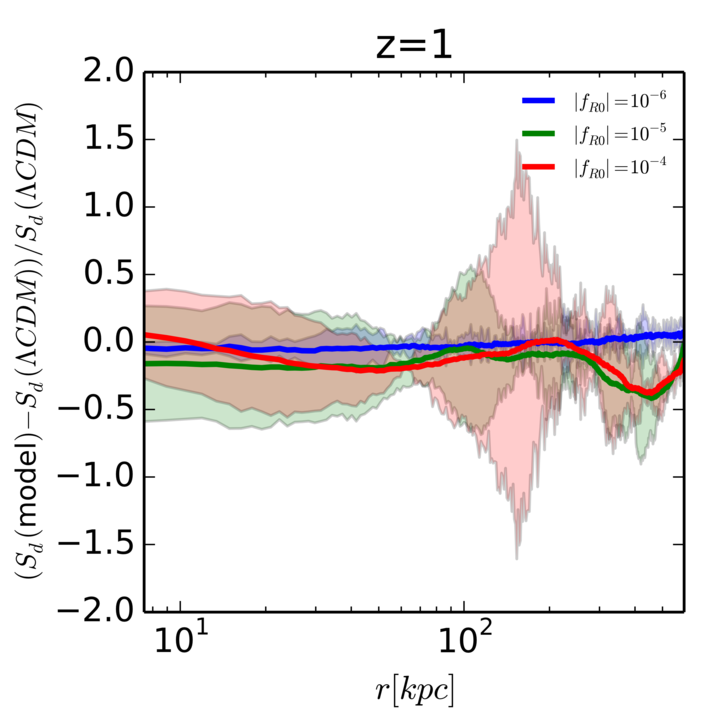} \\
\includegraphics[keepaspectratio,width=0.3\linewidth]{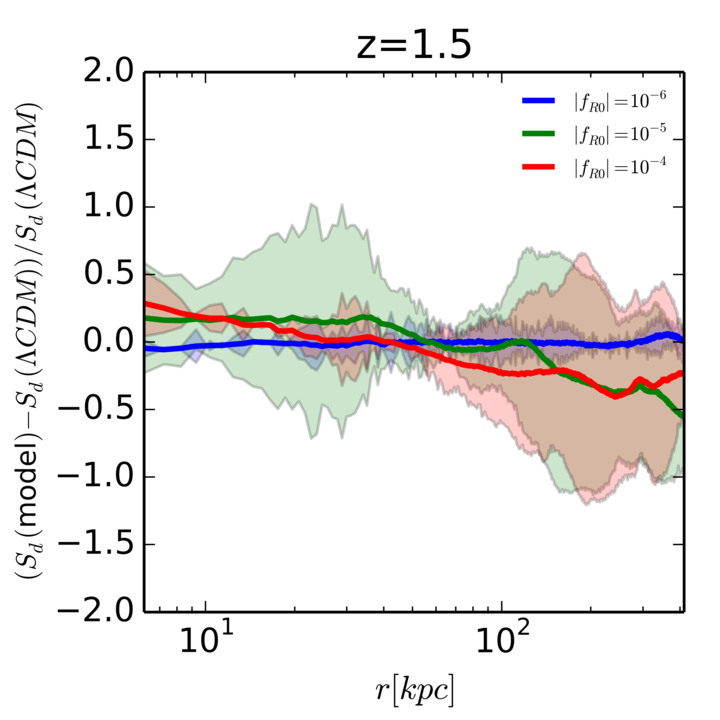} &
\includegraphics[keepaspectratio,width=0.3\linewidth]{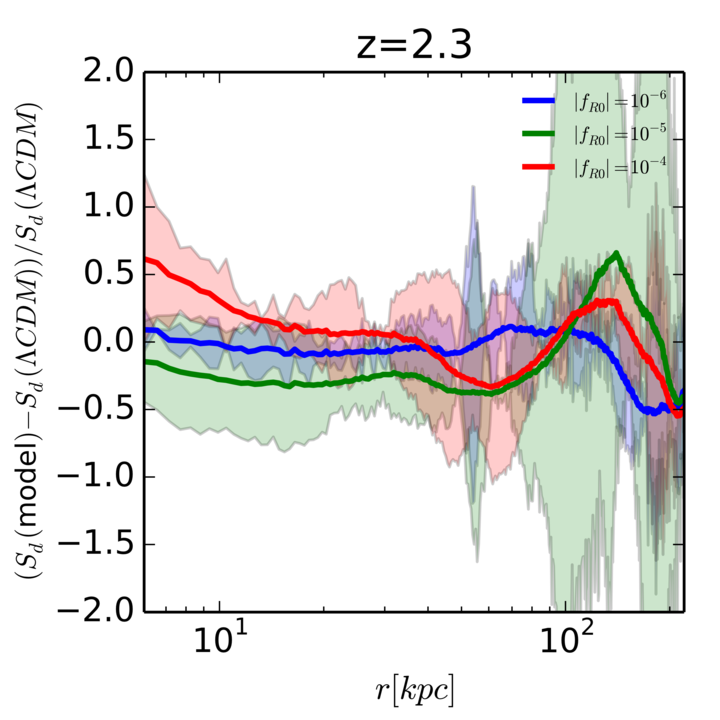} &
\includegraphics[keepaspectratio,width=0.3\linewidth]{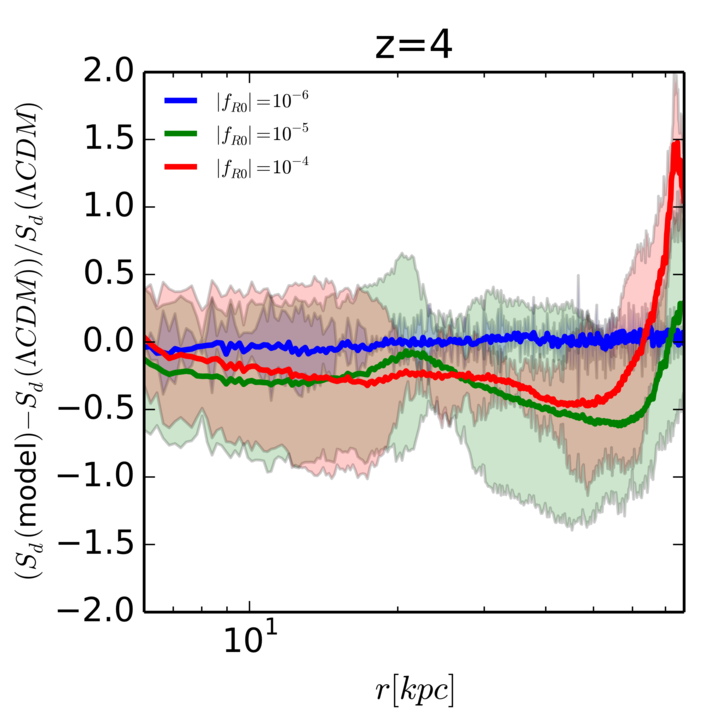} 
\end{array}$
\end{center}
\caption{Relative differences of cylindrical surface density profile to fiducial model. Error bars come from the variation of the quantity of 100 random lines of sight.}
\label{sddiff}
\end{figure*}

Considering the relative surface density profile in comparison to $\Lambda$CDM as depicted in \textbf{Figure \ref{sddiff}} we see that the distinguishability of the models has a stronger dependence on redshift and as with the velocity dispersion profiles the weak models are on average distinguishable from the strong models at all redshifts. In the centers of the profiles, the strong models show an enhancement of concentration at lower redshift, appearing as an enhancement of density relative to $\Lambda$CDM in the center of the halo, and a decrease in the outskirts. The higher the redshift the higher the concentration of the stronger $f(R)$ models relative to $\Lambda$CDM. The F4 model has higher concentration than the F5 model up to a redshift of 0.2 where the F5 model begins to have the greater concentration differential.

It is important to emphasize that for both \textbf{Figure \ref{sigmazdiff}} and in \textbf{Figure \ref{sddiff}} we note that the strong differences between the different models at high redshifts could be largely due to the different merger histories in these models, especially at the outskirts of the main halo. Therefore, although the difference can be substantial, one cannot claim that this can be used to distinguish one model from another universally, only that this discrimination is in principle possible for a given halo. A suite of simulations would need to be run at this high resolution, to determine the robustness of the distinguishability and disentangle the effect of merger history on the results.

\section{Lensing Signal}
\label{sec:lensing}
The relationship between the lensing potential and the matter density is found to be same in both GR and $f(R)$ scenarios with differences of order $|f_R|$ \citep{Zhao:2011bd}. $\Phi$ being the gravitational potential introduced above, $\Psi$ being the spatial curvature perturbation, the lensing potential is $(\Phi+\Psi)/2$ and we obtain for both scenarios:
\begin{equation}
	\label{eq:lensing}
	\nabla^2(\Psi + \Phi) = 8 \pi G a^2 \delta \rho_M
\end{equation}
We note that in general, $f(R)$ gravity will have a different matter overdensity evolution as a result of the modified solution, but the equation relating that overdensity to the lensing potential itself is the same as in GR.

In $f(R)$ gravity taking the modified Poisson equation \textbf{Equation \ref{eq:phi}}, we can represent it in a more familiar and simpler form by using $\delta \rho_{\rm{eff}}$, the perturbed total effective energy density, containing contributions from modifications to the Einstein tensor due to modified gravity and from matter.

\begin{equation}
    \nabla^2 \Phi=4 \pi G a^2 \delta \rho_{\rm{eff}}
\end{equation}

Thus for a given observational lensing signal, one can pick out a best fitting $\rho_M$ and see whether that distribution corresponds better to $f(R)$ gravity or unmodified general relativity. To examine lensing effects we can also separately consider the dynamical mass and the lensing mass, $M_D$ and $M_L$. The lensing mass is the true mass of an object, whereas the dynamical mass $M_D$ can be obtained from the Poisson equation (giving us $\delta \rho_{\rm{eff}}$) via
\begin{equation}
    \label{md}
    M_D \equiv \int a^2 \delta \rho_{eff} dV
\end{equation}
%TODO: detail how these are obtained from observations
Via an integration in spherical symmetry we obtain
\begin{equation}
    \label{mdr}
    M_D(r) = \frac{1}{G} r^2 d\phi(r)/dr
\end{equation}
$\Delta_M$ is defined as
\begin{equation}
    \label{deltamdef}
    \Delta_M \equiv M_D/M_L -1
\end{equation}
$\Delta_M$ ranges theoretically from 1/3 for unscreened galaxies to 0 for screened galaxies. Values above 1/3 are numerical artifacts.

We can consider two quantities to help in our analysis, namely $\Delta_M$ which can give insight into screening properties of particles, halos or subhalos, and $\Delta \Sigma$, the lensing signal which can give us insight into observables. Both can be measured directly from simulation.

The lensing signal is defined as:
 \begin{equation}
     \Delta \Sigma=\bar{\Sigma}(< R)-\Sigma(R)
  \end{equation} 
and can be computed from observations by inferring the lensing and the dynamical mass, respectively.

For our simulation we can compute the lensing signal directly as:
 \begin{equation}
     \bar{\Sigma}(< R)=\frac{1}{\pi R^2}\int_0^R\Sigma(R) 2 \pi R dR
 \end{equation}

\begin{figure}
\centering
\caption{Cylindrical lensing signal profile. Error bars come from the variation of the quantity of 100 random lines of sight.}
\label{lensingraw}
\begin{subfigure}{0.95\linewidth}
  \centering \includegraphics[keepaspectratio,width=0.95\linewidth]{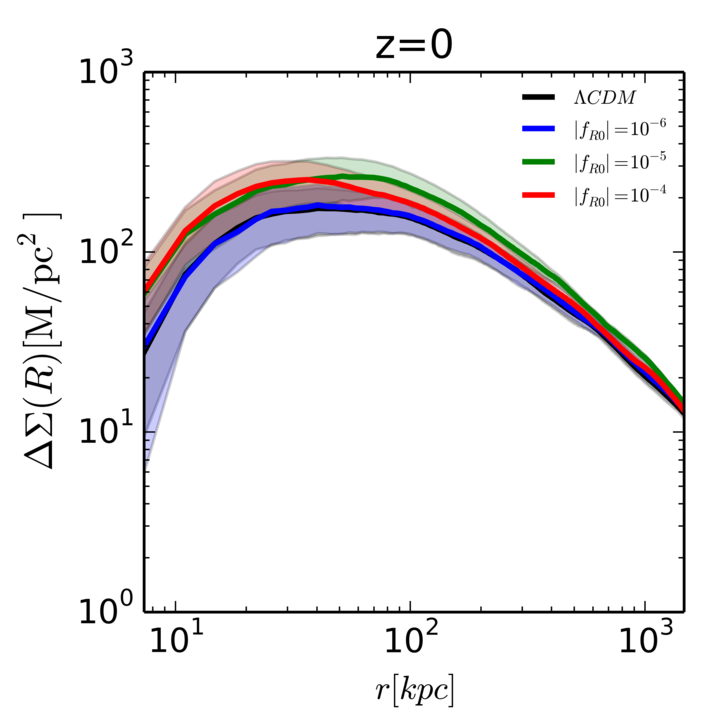}
\end{subfigure}
\begin{subfigure}{0.95\linewidth}
  \centering \includegraphics[keepaspectratio,width=0.95\linewidth]{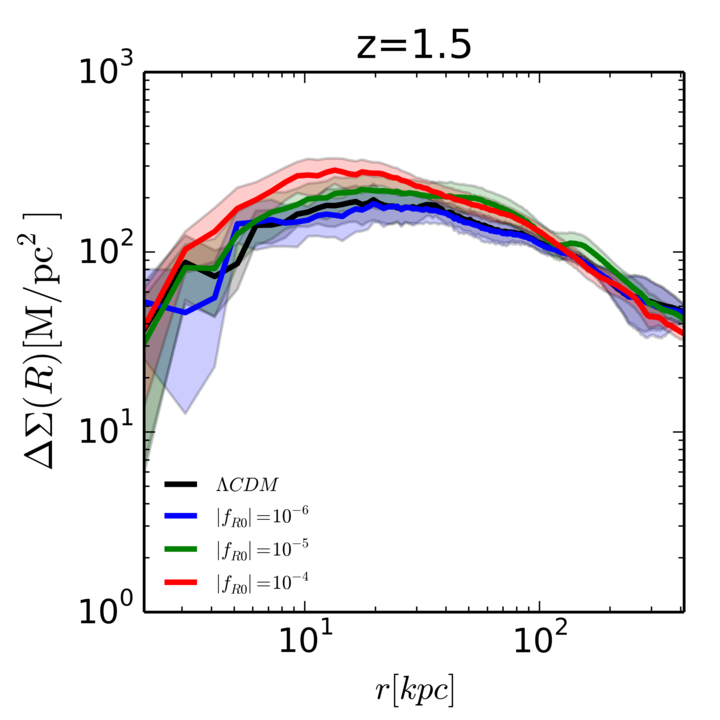}
\end{subfigure}
\end{figure}

We can see the results of the lensing signal for two redshifts in \textbf{Figure \ref{lensingraw}}. Here we see a clear peak in the lensing signal at an intermediate radius in all models. At outer radii and intermediate or high redshift the models are degenerate; the stronger models become distinguishable  only at lower radii and redshift. At higher redshift, for example, only the strongest model F4 is distinguishable from the other models and even then, only at lower radius. 

\begin{figure*}
\begin{center}$
\begin{array}{ccccc}
\includegraphics[keepaspectratio,width=0.3\linewidth]{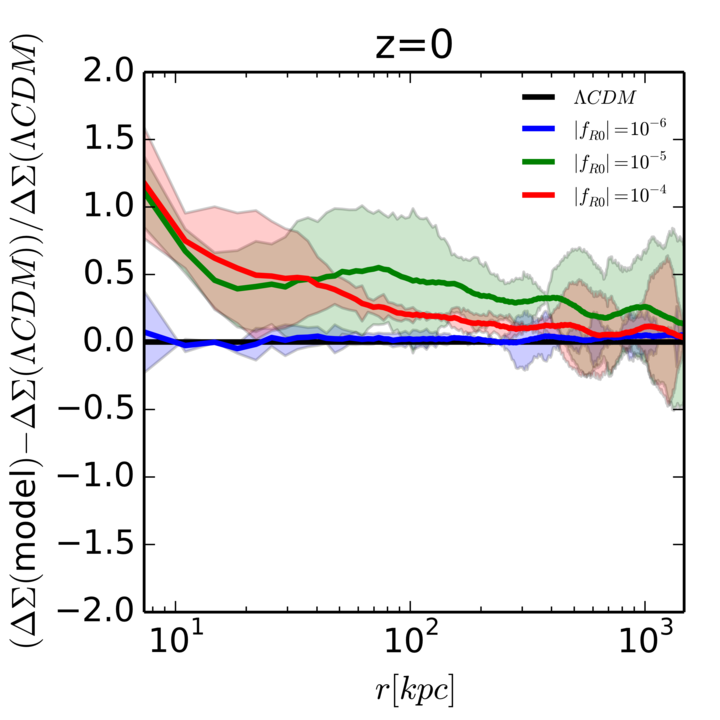} &
\includegraphics[keepaspectratio,width=0.3\linewidth]{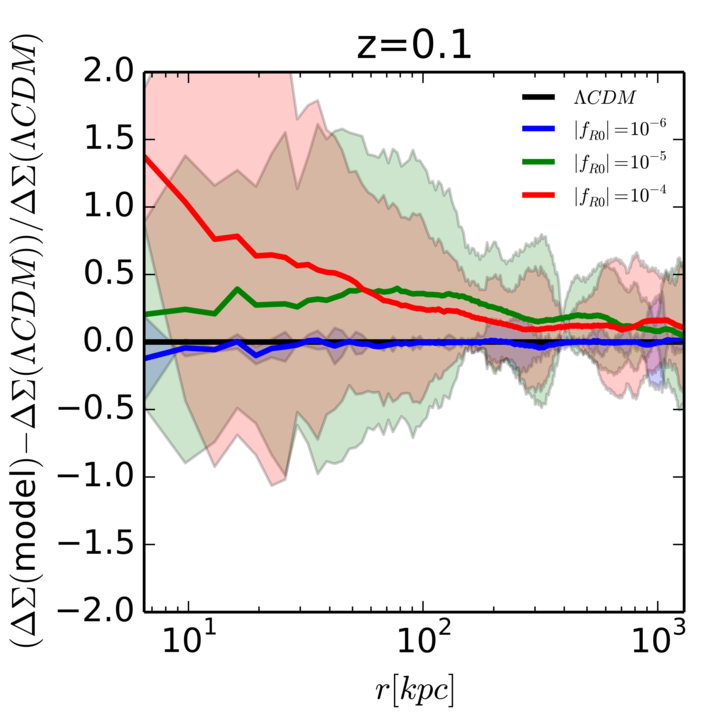} &
\includegraphics[keepaspectratio,width=0.3\linewidth]{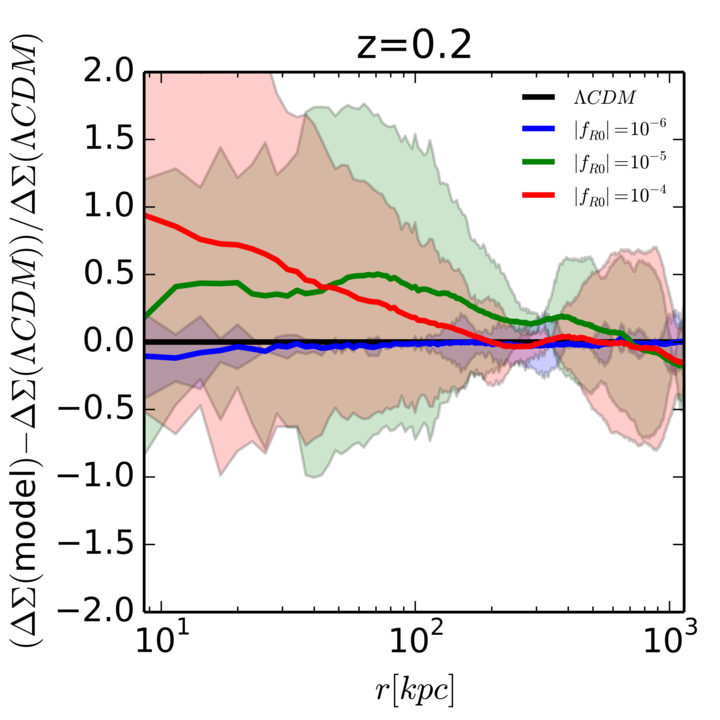} \\
\includegraphics[keepaspectratio,width=0.3\linewidth]{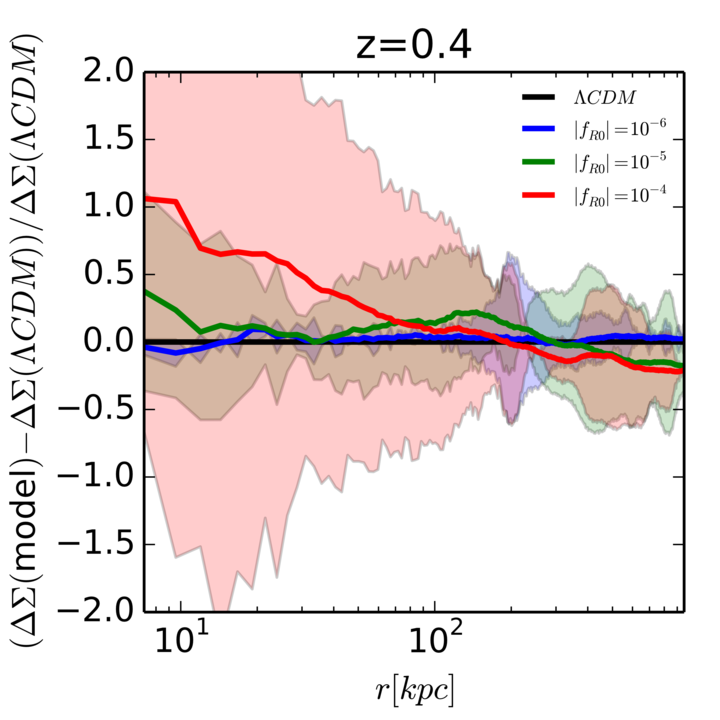} &
\includegraphics[keepaspectratio,width=0.3\linewidth]{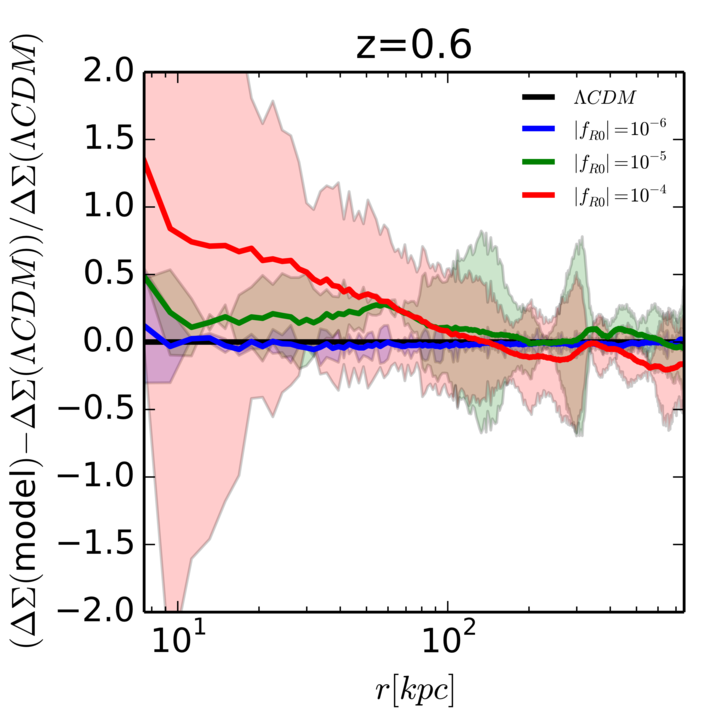} &
\includegraphics[keepaspectratio,width=0.3\linewidth]{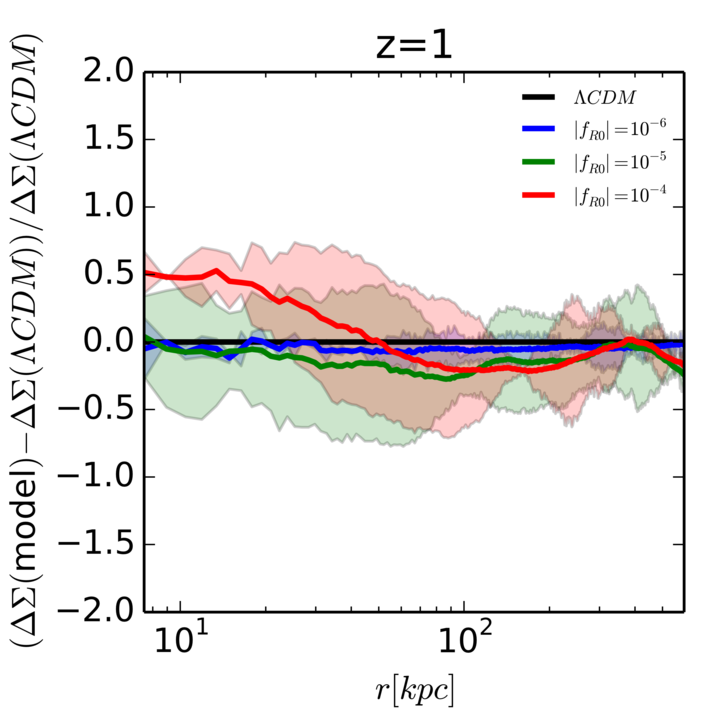} \\
\includegraphics[keepaspectratio,width=0.3\linewidth]{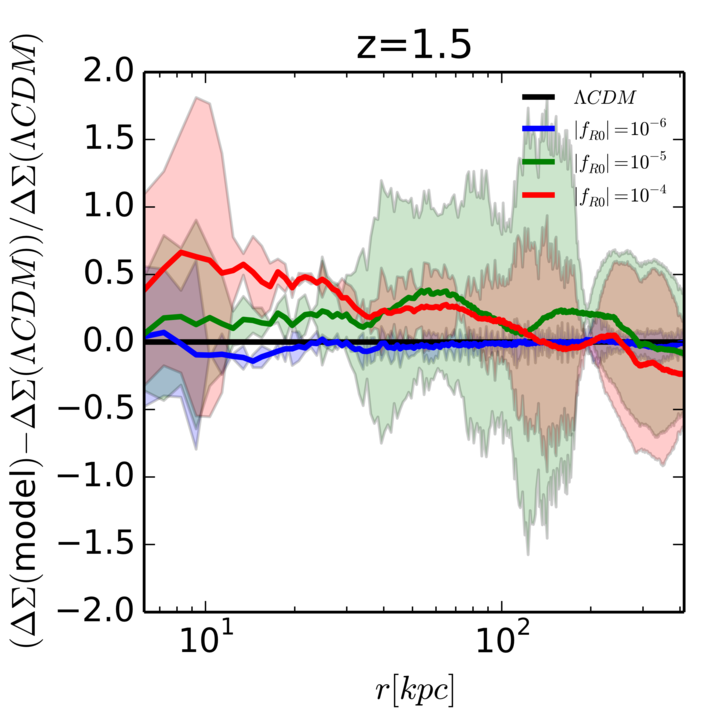} &
\includegraphics[keepaspectratio,width=0.3\linewidth]{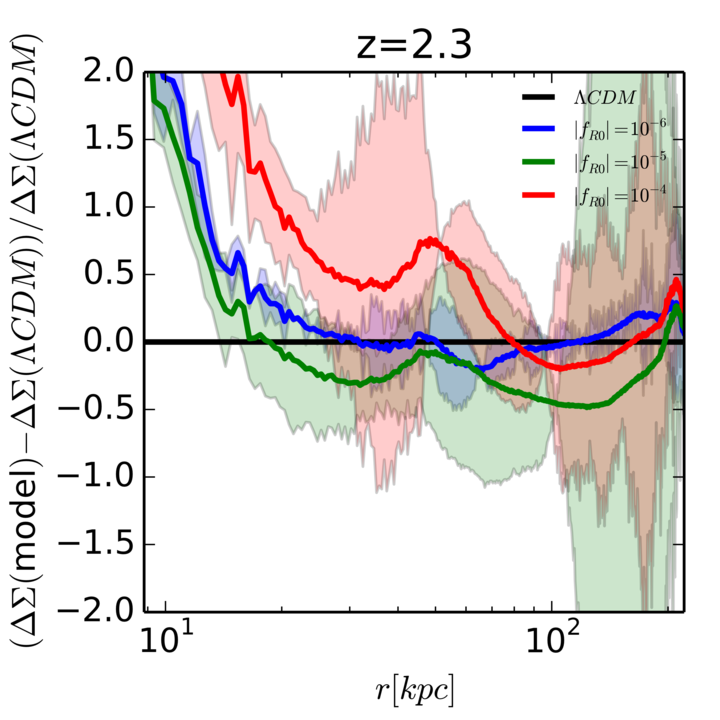} &
\includegraphics[keepaspectratio,width=0.3\linewidth]{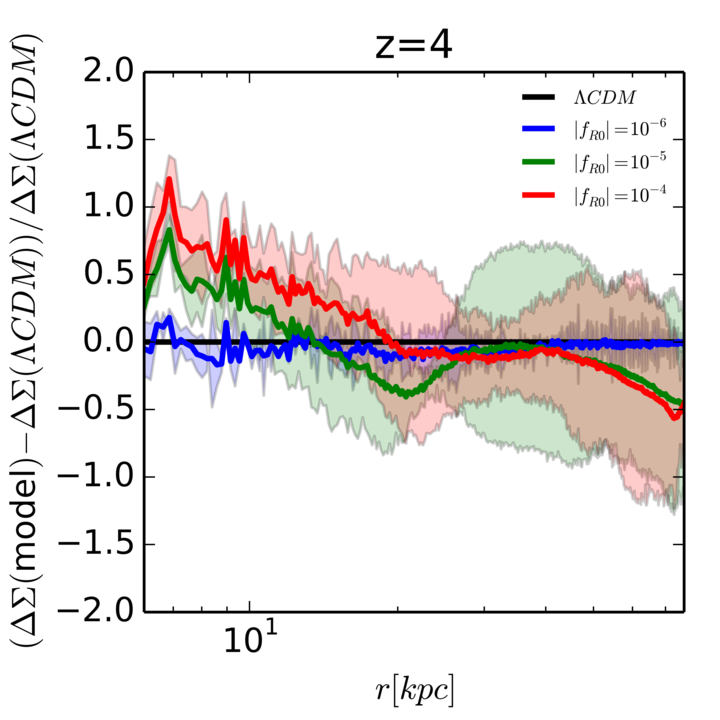} 
\end{array}$
\end{center}
\caption{Relative lensing signal. Error bars come from the variation of the quantity of 100 random lines of sight.}
\label{lensingdiff}
\end{figure*}

As with the surface density and velocity profiles, by examining the relative differences in the lensing signal between $f(R)$ models and $\Lambda$CDM we see the information conveyed by the lensing signal in \textbf{Figure \ref{lensingdiff}}. Here we see the strongest model becomes more distinguishable from $\Lambda$CDM with time in the inner parts of the halo. The slope of its relative difference to the $\Lambda$CDM model tends to be steeper near the center than that of the weaker F5 model, which tends to be closer to constant with radius. The same potential dependency for the lensing signal distinguishability on merger history applies as for the surface density distribution and the velocity dispersion results. The spike towards $r=0$ at $z=2.3$ is due to numerical centering issues as in the $f(R)$ gravity models two cores are seen.

\section{Physical Quantities}
\label{sec:physical}

We next plot $\nabla \Phi_{\rm{FR}}-\nabla \Phi_{\rm{GR}}$ vs. $\Phi_{\rm{GR}}$. For interpretation guidance, it is helpful to consider three regimes physically. In the chameleon regime  $\nabla\Phi_{\rm{FR}}-~\nabla\Phi_{\rm{GR}}=~0$, in the enhanced regime   $\nabla\Phi_{\rm{FR}}-~\nabla\Phi_{\rm{GR}}=~1/3\nabla \Phi_{\rm{GR}}$, whereas in the non-linear regime the value of this equation must be determined by solving the full coupled non-linear equations numerically.

\begin{figure*}
\begin{center}$
\begin{array}{cccc}
 F4 & F5 & F6 \\
\includegraphics[keepaspectratio,width=0.27\linewidth]{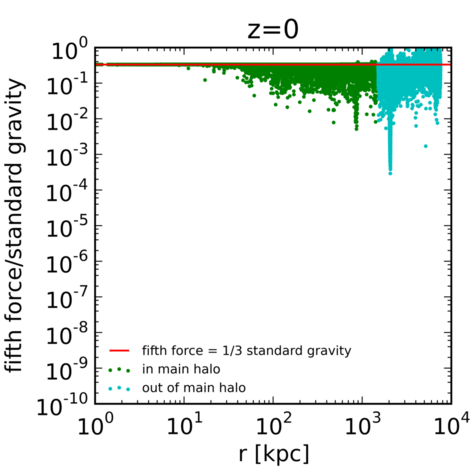} & \includegraphics[keepaspectratio,width=0.27\linewidth]{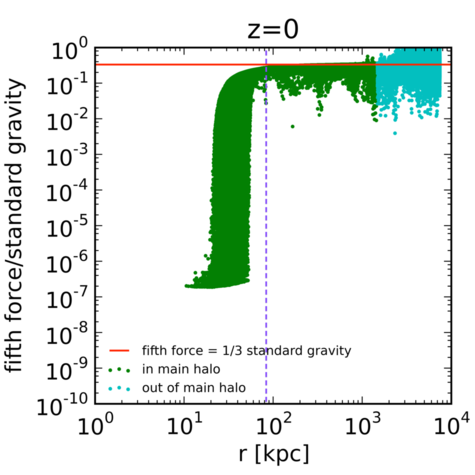} & \includegraphics[keepaspectratio,width=0.27\linewidth]{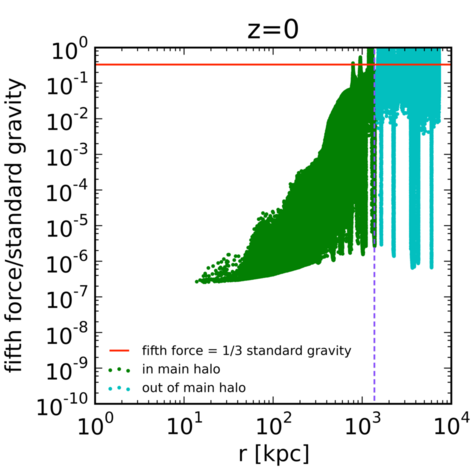} \\

\includegraphics[keepaspectratio,width=0.27\linewidth]{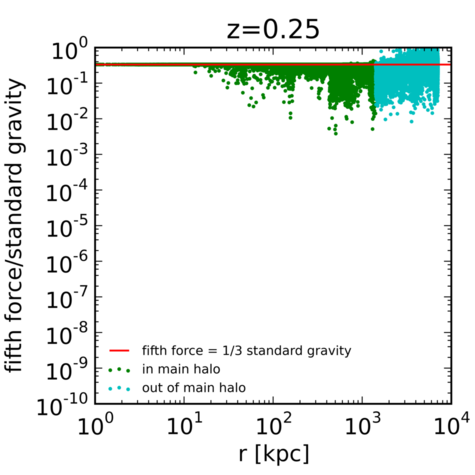} & \includegraphics[keepaspectratio,width=0.27\linewidth]{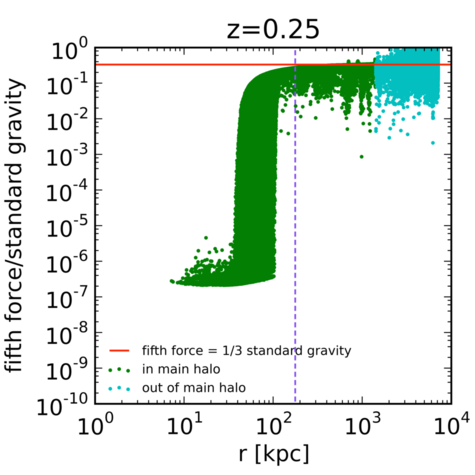} & \includegraphics[keepaspectratio,width=0.27\linewidth]{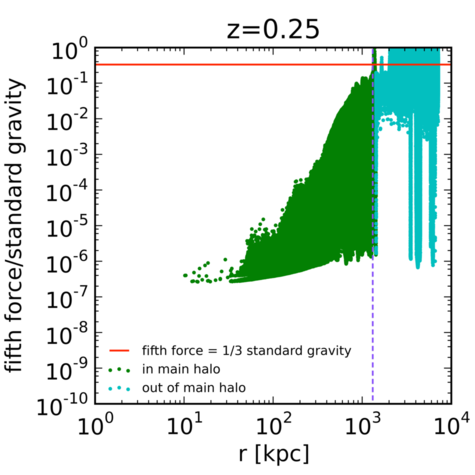} \\

\includegraphics[keepaspectratio,width=0.27\linewidth]{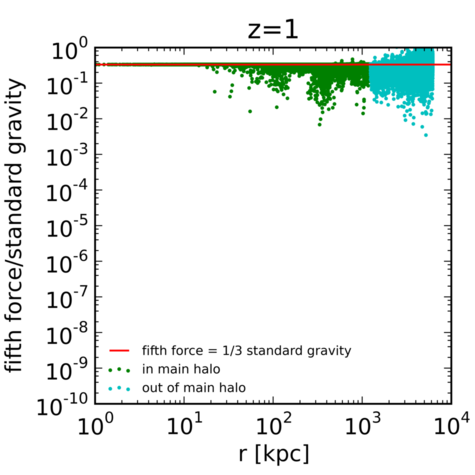} & \includegraphics[keepaspectratio,width=0.27\linewidth]{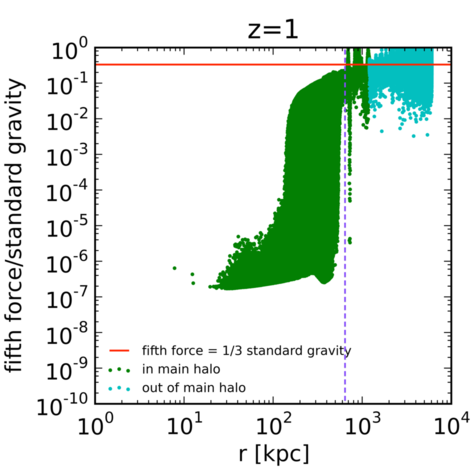}  & \includegraphics[keepaspectratio,width=0.27\linewidth]{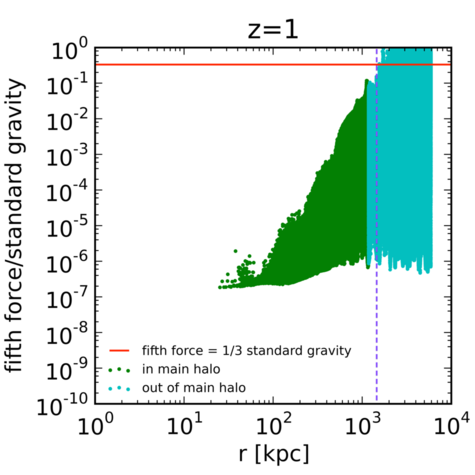} \\

\includegraphics[keepaspectratio,width=0.27\linewidth]{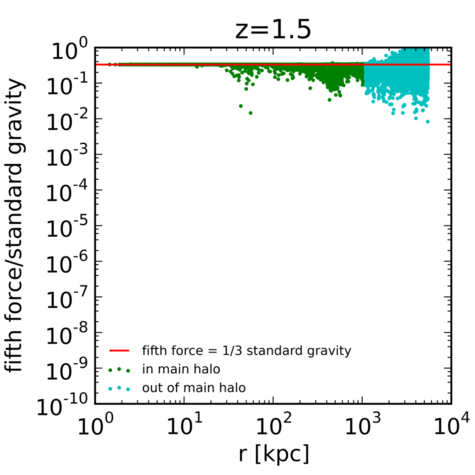} & \includegraphics[keepaspectratio,width=0.27\linewidth]{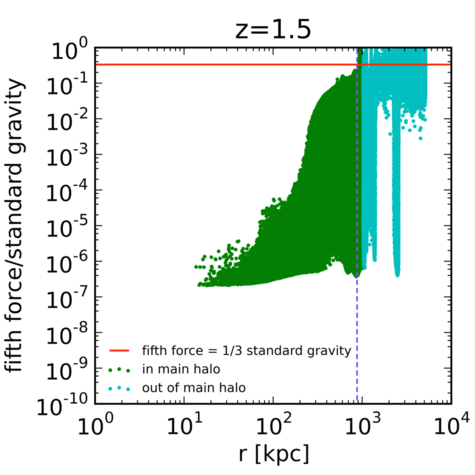} & \includegraphics[keepaspectratio,width=0.27\linewidth]{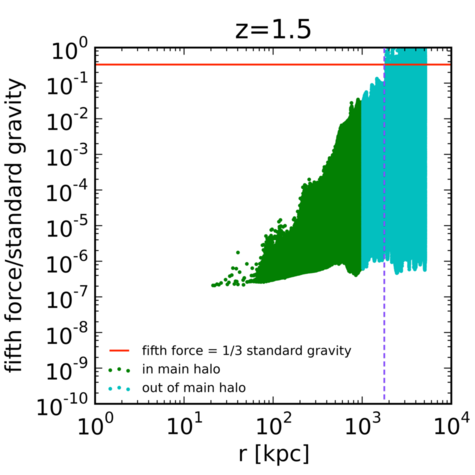} \\

\includegraphics[keepaspectratio,width=0.27\linewidth]{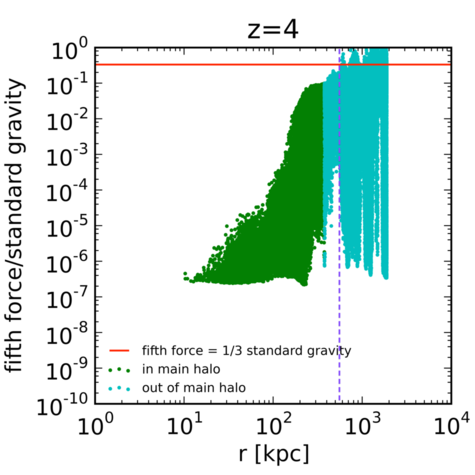} & \includegraphics[keepaspectratio,width=0.27\linewidth]{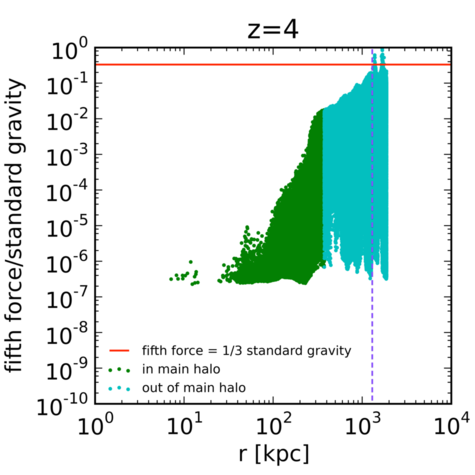} & \includegraphics[keepaspectratio,width=0.27\linewidth]{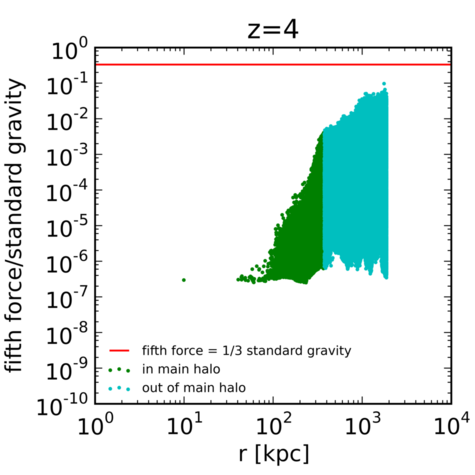} \\

\end{array}$
\end{center}
\caption{Force ratio for particles versus radius. Particles are colored by whether they belong to the main halo (green) or lie outside the main halo (cyan). Horizontal purple dashed lines correspond to visual heuristic marking radius at where transition from unscreened to screened begins.}
\label{ratioforces}
\end{figure*}

We first separated particles by whether they were residing within the main halo, and by radius from the center of the main halo. \textbf{Figure \ref{ratioforces}} depicts this result and shows the ratio of the fifth force to that of standard gravity, which approaches zero in the screened regions and $1/3$ in the regions of enhanced gravity, again values above $1/3$ are to be considered numerical artifacts, with in general symmetric scatter about $1/3$.

First considering the particles within the main halo, for the F6 model we see chameleon effect dominates at all redshifts and only by $z=0$ do we start to see some enhancement. For the F5 model in the main halo more enhancement than F6 in main halo is seen, and this enhancement begins at $z=1.5$. Overall the F5 model has a mix of chameleon effect and enhancement. For the strongest F4 model in the main halo we see enhancement dominates by $z=1.5$, and it is only prior to that the chameleon effect is important. On the whole, the chameleon effect is important in the main halo, as we would expect and dominates at high redshift. Enhancement does become important in each model even in the main halo. The redshift at which it becomes important is lowest for the weakest model and highest for the strongest model likewise as we would expect.
    
Next, considering the particles outside of the main halo we see a mix of effects. For the F6 model, a larger region of enhancement begins around $z=2.3$. For the F5 model the chameleon effect is no longer important by $z=1$; from $z=1.5$ to $z=4$ a mix of enhancement and the chameleon effect is seen, while at high $z$ the chameleon effect dominates. For the F4 model enhancement already occurs at $z=4$, dominates after that, and the chameleon effect in this model is only important at very high $z$.

Comparing the two, whereas the enhancement region was almost entirely absent at all redshifts for F6 within the main halo, outside there is a mix. Likewise for F5 the chameleon effect is only important outside of the main halo at very high redshift, with the enhancement region quickly dominating whereas inside of the main halo there is a mix of effects. The difference between the two regimes is less apparent with the strongest F4 model, as both particles in and out of the main halo begin to feel enhancement early on, with onset of importance for particles in the main halo being simply later in time. Thus dividing particles on the basis of environment in this manner is most effective for the weakest models. This is an important result as it means that the environmental differences between the different models can show up more saliently at lower redshift for weaker, not for stronger models, for which by low redshift the macro environmental differences have disappeared.

We see similar pattern as a function of radius present in each simulation, only the redshift at which a certain profile is seen changing depending on the strength of the model. We identify the transition redshift at which any particles within the main halo become subject to the full fifth force enhancement in each model as a critical stage for each model. This is $z \approx 4$ for F4 $z\approx 1.5$ for F5, and $z\approx 0$ for F6. Likewise we identify the redshift for which all particles in the main halo become subject to the full fifth force enhancement. This quite clearly occurs before $z=1.5$ for the F4 model, but is only on its way to occurring (the outer part of halo fully subject to enhancements, but not the inner portion) for F5. For F6 we would have to simulate further in the future to find the redshift at which this occurs, but it is quite clearly also on this evolutionary path, with the outskirts of the halo being effected by modified gravity at this stage.

As such we could propose a cutoff radius for $f(R)$ models which is a function of both redshift and model for the main halo considered. Within this cutoff radius, the chameleon effect is present, and outside of this radius the enhancement is in effect. This cutoff radius can be read off of \textbf{Figure \ref{ratioforces}}, which are shown as the vertical dashed lines.

To show more accurate environmental dependence of such effects, we next separated particles by their environment, starting with whether they reside inside or outside of the main halo. We used \texttt{ROCKSTAR} \citep{2013ApJ...762..109B} to identify halos and subhalos and within our simulation.  \texttt{ROCKSTAR} is a 7D (temporal) optionally 6D structure and substructure phase space finder. \texttt{ROCKSTAR} uses Friends-of-Friends groups with a large linking length as a parameter to divide the volume in 3D. Next for each group it is ensured that 70\% of its particles are linked in subgroups which implies an adaptive phase-space (6D) metric. This procedure is recursively applied with a final step to assemble the seed halos in their densest subgroups by assigning each particle to the group closest to it in phase space. Finally, an unbinding procedure is used. We deploy \texttt{ROCKSTAR} in its 6D implementation to form a full halo and subhalo catalog at each redshift in each model. 

\begin{figure*}
\begin{center}$
\begin{array}{cccc}
 z & F4 & F5 & F6 \\
 0 & \includegraphics[keepaspectratio,width=0.26\linewidth]{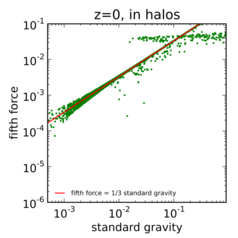} & \includegraphics[keepaspectratio,width=0.26\linewidth]{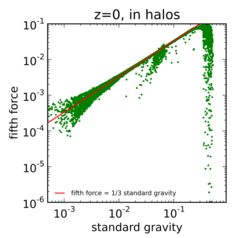} & \includegraphics[keepaspectratio,width=0.26\linewidth]{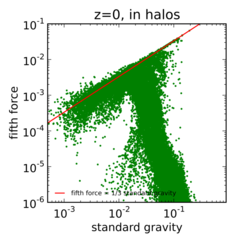} \\
 0.25 & \includegraphics[keepaspectratio,width=0.26\linewidth]{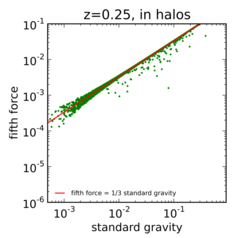} & \includegraphics[keepaspectratio,width=0.26\linewidth]{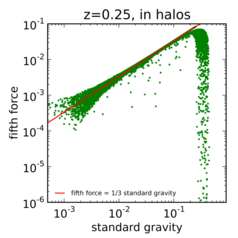} & \includegraphics[keepaspectratio,width=0.26\linewidth]{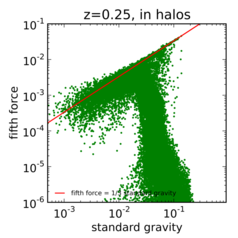} \\
 1.5 & \includegraphics[keepaspectratio,width=0.26\linewidth]{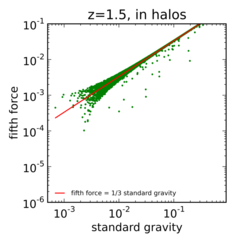} & \includegraphics[keepaspectratio,width=0.26\linewidth]{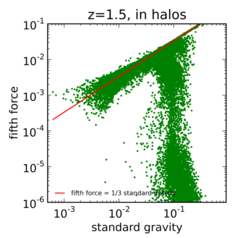} & \includegraphics[keepaspectratio,width=0.26\linewidth]{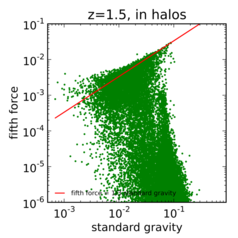} \\
 4 & \includegraphics[keepaspectratio,width=0.26\linewidth]{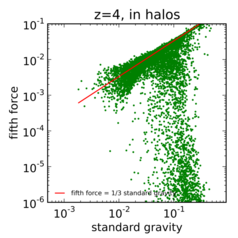} & \includegraphics[keepaspectratio,width=0.26\linewidth]{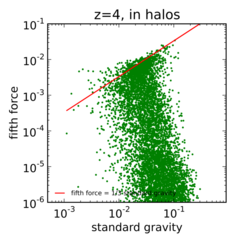} & \includegraphics[keepaspectratio,width=0.26\linewidth]{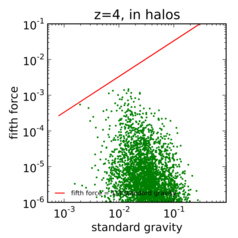} \\
\end{array}$
\end{center}
\caption{Force for particles within halos or subhalos.}
\label{forcesinhalos}
\end{figure*}

\begin{figure*}
 \begin{center}$
 \begin{array}{cccc}
 z & F4 & F5 & F6 \\
 0 & \includegraphics[keepaspectratio,width=0.26\linewidth]{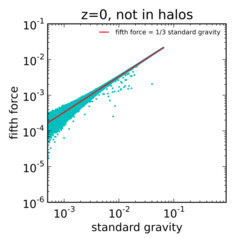} & \includegraphics[keepaspectratio,width=0.26\linewidth]{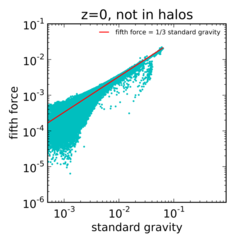} & \includegraphics[keepaspectratio,width=0.26\linewidth]{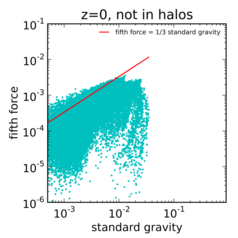} \\
 0.25 & \includegraphics[keepaspectratio,width=0.26\linewidth]{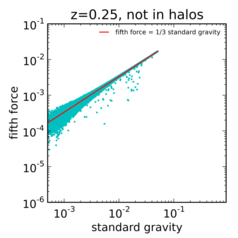} & \includegraphics[keepaspectratio,width=0.26\linewidth]{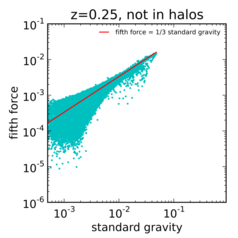} & \includegraphics[keepaspectratio,width=0.26\linewidth]{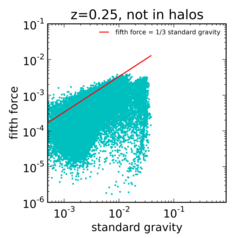} \\
 1.5 &  \includegraphics[keepaspectratio,width=0.26\linewidth]{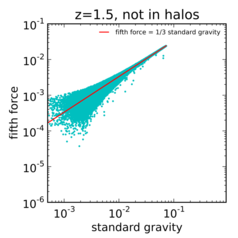} & \includegraphics[keepaspectratio,width=0.26\linewidth]{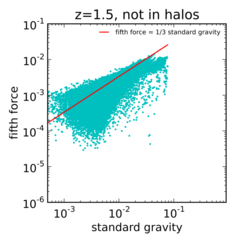} & \includegraphics[keepaspectratio,width=0.26\linewidth]{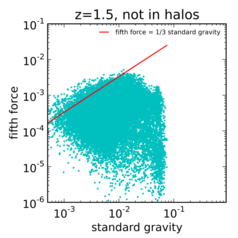} \\
 
 4 & 
\includegraphics[keepaspectratio,width=0.26\linewidth]{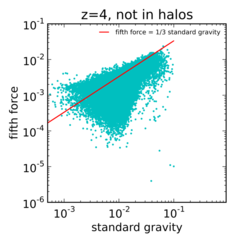} & \includegraphics[keepaspectratio,width=0.26\linewidth]{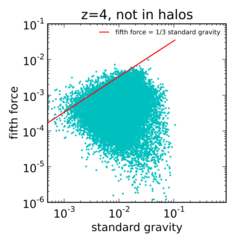} & \includegraphics[keepaspectratio,width=0.26\linewidth]{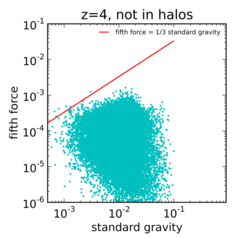} \\
\end{array}$
\end{center}
 \caption{Force for particles outside of halos or subhalos.}
 \label{forcesnotinhalos}
 \end{figure*}

We see in \textbf{Figure \ref{forcesinhalos}} and \textbf{Figure \ref{forcesnotinhalos}} that particles within halos show a mixture of enhancement and screening effects, but particles outside of halos are almost entirely unscreened independent of the model and redshift under consideration, with only the weakest F6 model showing some outside of halo screening at $z \geq 1.5$.

\section{Screening of satellites}

\cite{Cabre:2012cm} use N-body simulations by \cite{Zhao:2011bd} to test and calibrate methods to determine the level of screening in galaxy catalogs. Due to the chameleon mechanism GR should be restored in massive halos or high density environments. Observational constraints in the solar system rule out the presence of modified gravity effects in galaxy halos with masses higher than the Milky Way $\approx 10^{12} M_\odot$, however field dwarf galaxies, whose Newtonian potential is at least an order of magnitude smaller than their hosts, may be unscreened and be impacted by the extra force. Thus to constrain $f(R)$ gravity such galaxies are particularly valuable. In particular, as the tests using distance indicators as sensitive probes rely on observations within hundreds of Mpc, classifying galaxies as being screened or unscreened within this region is especially important.

Our simulations, performed at higher resolution with a fully AMR code will be informative in this respect, with the particular ability to discriminate at the satellite level whether the satellite is screened or unscreened. In particular, \cite{Cabre:2012cm} use only $256^3$ particles in a box of $64$ Mpc/h, where we have much higher resolution. Their minimum particle mass $10^9 M_\odot h^{-1}$ whereas ours is $3.6 \times 10^7 M_\odot h^{-1}$, two orders of magnitude smaller.  In this manner we can make comments on halos which are field dwarf analogues for our cluster, and compare and confirm the accuracy of the \cite{Cabre:2012cm} predictions. We review our method for identifying satellites of the main halo in our simulations, the techniques deployed by \cite{Cabre:2012cm}, and their application to our simulations.

\begin{figure*}
\begin{center}$
    \begin{array}{cc}
\begin{array}{ccc}
F4 & F5 & F6 \\    
\includegraphics[keepaspectratio,width=0.3\linewidth,height=0.14\textheight]{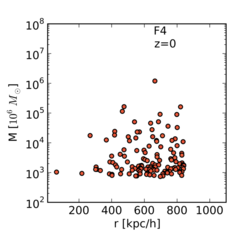} &
\includegraphics[keepaspectratio,width=0.3\linewidth,height=0.14\textheight]{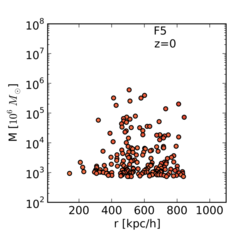} &
\includegraphics[keepaspectratio,width=0.3\linewidth,height=0.14\textheight]{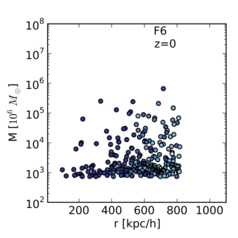} 
\\
\includegraphics[keepaspectratio,width=0.3\linewidth,height=0.14\textheight]{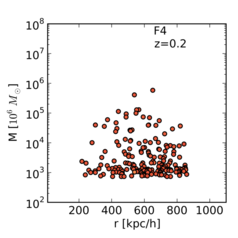} &
\includegraphics[keepaspectratio,width=0.3\linewidth,height=0.14\textheight]{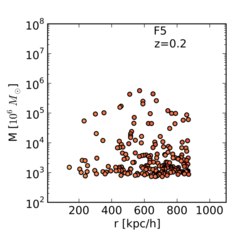} &
\includegraphics[keepaspectratio,width=0.3\linewidth,height=0.14\textheight]{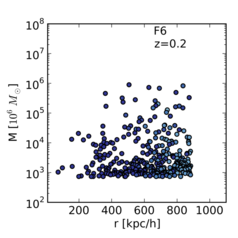} 
\\
\includegraphics[keepaspectratio,width=0.3\linewidth,height=0.14\textheight]{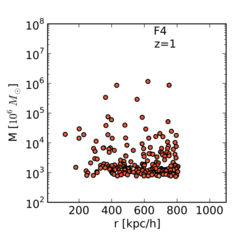} &
\includegraphics[keepaspectratio,width=0.3\linewidth,height=0.14\textheight]{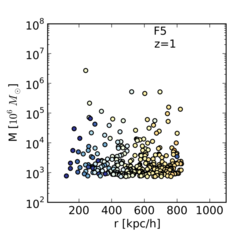} &
\includegraphics[keepaspectratio,width=0.3\linewidth,height=0.14\textheight]{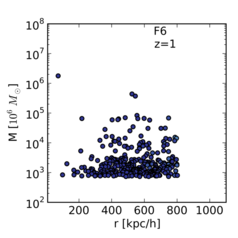} 
\\
\includegraphics[keepaspectratio,width=0.3\linewidth,height=0.14\textheight]{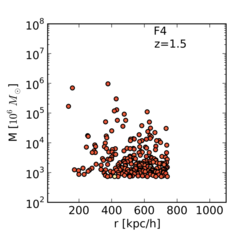} &
\includegraphics[keepaspectratio,width=0.3\linewidth,height=0.14\textheight]{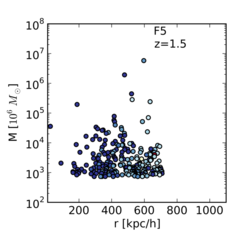} &
\includegraphics[keepaspectratio,width=0.3\linewidth,height=0.14\textheight]{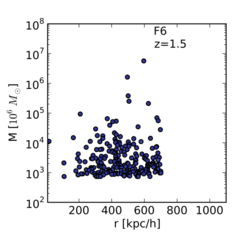} 
\\
\includegraphics[keepaspectratio,width=0.3\linewidth,height=0.14\textheight]{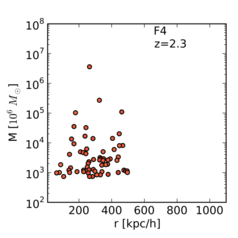} &
\includegraphics[keepaspectratio,width=0.3\linewidth,height=0.14\textheight]{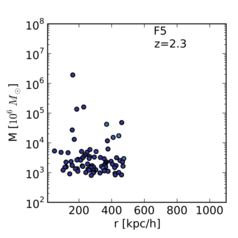} &
\includegraphics[keepaspectratio,width=0.3\linewidth,height=0.14\textheight]{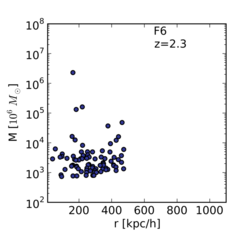} 
\\
\includegraphics[keepaspectratio,width=0.3\linewidth,height=0.14\textheight]{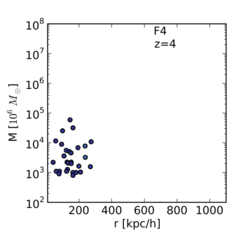} &
\includegraphics[keepaspectratio,width=0.3\linewidth,height=0.14\textheight]{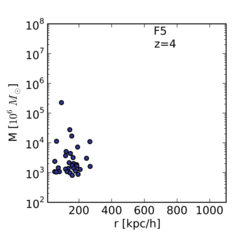} & 
\includegraphics[keepaspectratio,width=0.3\linewidth,height=0.14\textheight]{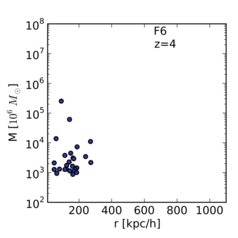} 
\end{array}
& \specialcell{\includegraphics[keepaspectratio,height=0.14\textheight]{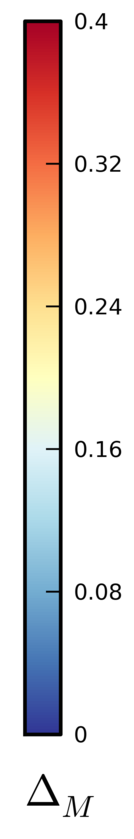}}
\end{array}$
\end{center}
\caption{$\Delta_M$ as a function of radius and host halo mass. }
\label{deltam}
\end{figure*}

We again use \texttt{ROCKSTAR} to identify subhalos of the main halo of our simulations as satellites to investigate the physical effects of modified gravity on satellites. \textbf{Figure \ref{deltam}} depicts these results of $\Delta_M$ as a function of radius and satellite mass as measured within our simulation. Here we see that the transition epochs from screened to unscreened occur for satellites at the same epoch as for particles within the main halo as a whole. More specifically, we see that environment has a strong effect on how this transition proceeds, with radius of the satellite from the main halo the predominant component. This can be seen most clearly in the $z=1.5$ and $z=1$ subplots of the F5 model, where satellites closest to the main halo remain screened while satellites in the outskirts transition to unscreened. This is the predominant but not the only effect, as there are some satellites residing in the outer part that remain screened and vis versa. 

The result that the satellite halos switch from screened to un-screened at about the same time as the particles is not particularly surprising, because the particles inside the main halo contain the particles in the satellites. This result is an important one in the context of lower resolution simulations however, as it shows that knowing the screening of particles implies a knowledge of the screening of the satellites. Low resolution simulations are unable to resolve the satellite halos very well, but this finding shows that they can still be used to inform whether subhalos are screened or not in the main halo.

While $\Delta_M$ is directly measurable in simulation, in practice it can be difficult to measure the lensing mass, which is necessary to compute this theoretical quantity. \cite{Cabre:2012cm} deployed and tested two criteria for environmental based screening, both readily inferred from observational data, designed to be deployed on galaxy catalogs from observations. We deploy a similar set of criterion in our high resolution simulations to test their applicability at the subhalo level and to try to tease out potential additional influences of the environment on the screening of a satellite.

The first, and simpler hypothesis they deploy is that a galaxy cannot be effected  by a nearby galaxy's fifth force if it is beyond the Compton wavelength from the outer boundary of that galaxy. Thus  \cite{Cabre:2012cm} devise a criterion that a test galaxy $i$ will only feel a fifth force within a distance of
\begin{equation}
    \lambda_C+r_i
\end{equation}
assuming the galaxy is not self-screened. They find that this simple classification scheme works in simulation.

Formalizing this criterion per \cite{Cabre:2012cm}, a halo is self-screened at $z=0$ if it satisfies
\begin{equation}
    \label{selfscreenedcabre}
    \frac{|\phi_{\rm{int}}|}{c^2} > \frac{3}{2} |f_{R0}|
\end{equation}
or
\begin{equation}
    \label{environmentscreenedcabrre}
    \frac{|\phi_{\rm{ext}}|}{c^2} > \frac{3}{2} |f_{R0}|
\end{equation}
This is motivated by the fact that GR is recovered if the model parameter $f_{R0}$ is less than $\frac{2}{3} \frac{|\phi_N|}{c^2}$ where $\phi_N$ is the Newtonian potential; this recovery is due to the chameleon effect \citep{Hu:2007tv}. 

Here the internal screening is computed via:
\begin{equation}
    |\phi_{\rm{int}}|=\frac{GM}{r_{\rm{vir}}}
\end{equation}

Motivated by the fact the range of the fifth force is finite we likewise define:
\begin{equation}
    |\phi_{\rm{ext}}|=\sum_{d_i<\lambda_C+r_{\rm{vir,i}}} \frac{GM_i}{d_i}
\end{equation}
where $d_i$ is the distance to a neighbor halo with mass $M_i$ and radius $r_{\rm{vir,i}}$.

We extend their analysis beyond $z=0$ by computing the background value of $f_R$ as a function of redshift. Noting that $\bar{R}=3m^2(a^{-3}+\frac{2}{3}\frac{c_1}{c_2})$, $f_{\bar{R}}\approx -n \frac{c_1}{c_2^2}(\frac{M^2}{-\bar{R}})^{1+n}$ and $\frac{c_1}{c_2^2}=\frac{1}{n}\left[ 3(1+4\frac{\Omega_\Lambda}{\Omega_m})\right]^{1+n} |f_{R0}|$, for our chosen value of $n$ and model (giving us $f_{R0}$), we can solve for $f_{\bar{R}z}$:
\begin{equation}
    f_{\bar{R}z}=(\frac{1+4\frac{\Omega_\Lambda}{\Omega_m}}{a^{-3}+4\frac{\Omega_\Lambda}{\Omega_m}})^2 f_{R0}
\end{equation}
and modify the self screened and environmentally screened criterion to read:
\begin{equation}
    \label{selfscreened}
    \frac{|\phi_{\rm{int}}|}{c^2} > \frac{3}{2} |f_{\bar{R}z}|
\end{equation}
or
\begin{equation}
    \label{environmentscreened}
    \frac{|\phi_{\rm{ext}}|}{c^2} > \frac{3}{2} |f_{\bar{R}z}|
\end{equation}

\begin{figure*}
\begin{center}$
\begin{array}{ccccc}
\includegraphics[keepaspectratio,width=0.3\linewidth]{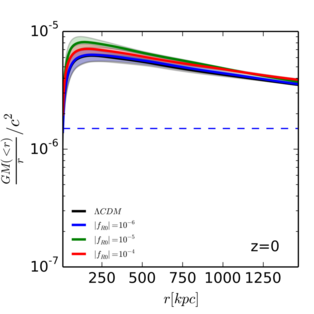} &
\includegraphics[keepaspectratio,width=0.3\linewidth]{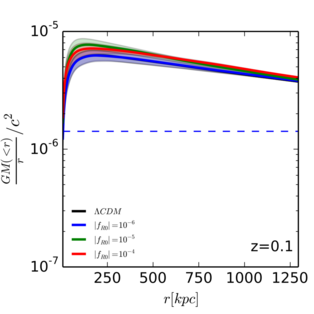} &
\includegraphics[keepaspectratio,width=0.3\linewidth]{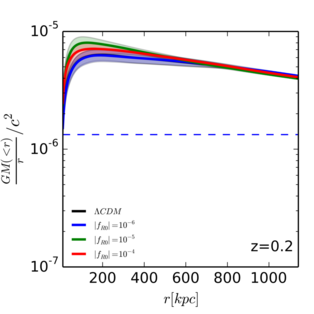} \\
\includegraphics[keepaspectratio,width=0.3\linewidth]{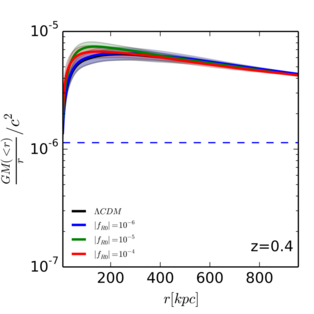} &
\includegraphics[keepaspectratio,width=0.3\linewidth]{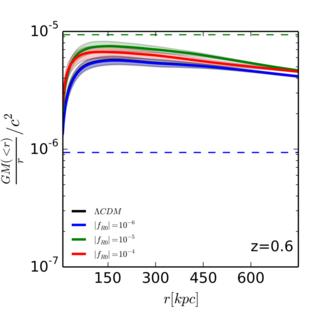} &
\includegraphics[keepaspectratio,width=0.3\linewidth]{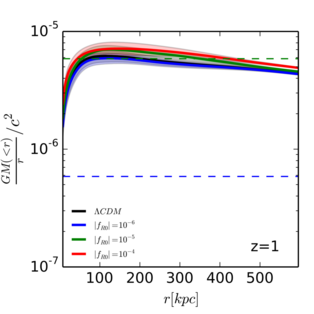} \\
\includegraphics[keepaspectratio,width=0.3\linewidth]{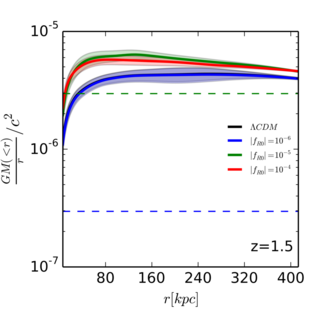} &
\includegraphics[keepaspectratio,width=0.3\linewidth]{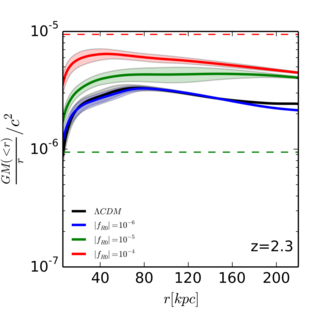} &
\includegraphics[keepaspectratio,width=0.3\linewidth]{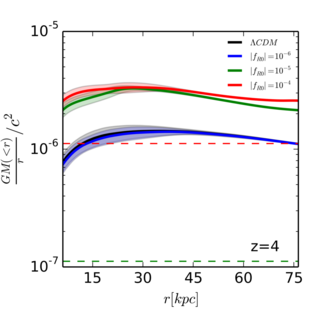} 
\end{array}$
\end{center}
\caption{Level of screening for the main halo. Dashed lines are colored according their respective models and correspond to $\frac{3}{2}|f_{\bar{R}z}|$. The solid profiles which reside above their corresponding dashed line are predominantly screened at that redshift, those which reside below are predominantly unscreened/enhanced at that redshift and those that intersect correspond to a transition between the two regimes. }
\label{screening}
\end{figure*}

In \textbf{Figure \ref{screening}} we see the behavior of $|\phi_{\rm{int}}|(r)=\frac{GM(<r)}{r}$ as compared to $\frac{3}{2}|f_{\bar{R}z|}$ to test the hypothesis of being able to use the latter as a measure of screening. For the F4 model we can see only at $z\approx4$ is the level of background screening on the order of the halo potential, so for this model this confirms theoretically that only at high redshift $z\geq 4$ do we expect any level of screening in the main halo. For F5 we see screening to occur at $z \geq 2.3$ and a transition region to take place in between $z=1.5$ and $z=1$. By $z=0.6$ the  halo is unscreened. For the F6 model, only at $z=0$ does there begin to be a hint of a transition from screened to unscreened. This confirms what we see in \textbf{Figure \ref{ratioforces}} as to the physical values of the epoch of transition. Thus the relationship between the halo potential and $\frac{3}{2}f_{\bar{R}z}$ can be used as a proxy for screening at a given epoch.

\cite{Cabre:2012cm}  find that a simple cut in the $\log(|\phi_{ext}|)/c^2$ vs $\log(M_{dyn})$ plane can discriminate between screened and unscreened halos. They predict that only the largest halos in F5 (above $10^{14} M_\odot h^{-1}$) or those with an especially high external potential will be screened in F5 at $z=0$, so a main halo on the order of our simulated halo is expected to be fully unscreened at $z=0$. For F6, halos on above $10^{13} M_\odot h^{-1}$ or with an especially high external potential will either be screened or transitioning to being unscreened at $z=0$. This behavior is exactly what we see in \textbf{Figure \ref{deltam}} where the satellites of our main halo in F5 are fully unscreened and those in F6 are undergoing a transition for at $z=0$. 

We can test the validity of this scheme as applied to subhalos in our simulation by plotting likewise $\log(|\phi_{ext}|)/c^2$ vs $\log(M_{dyn})$ and coloring by the relative difference $\Delta_M$ between the dynamical and lensing masses of the same object given by \textbf{Equation \ref{deltamdef}}, to quantify the effects of modified gravity within an object. 

\begin{figure*}
\begin{center}$
\begin{array}{cc}
\begin{array}{ccc}
F4 & F5 & F6 \\    
\includegraphics[keepaspectratio,width=0.3\linewidth,height=0.14\textheight]{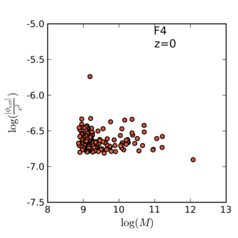} &
\includegraphics[keepaspectratio,width=0.3\linewidth,height=0.14\textheight]{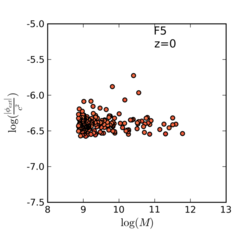} &
\includegraphics[keepaspectratio,width=0.3\linewidth,height=0.14\textheight]{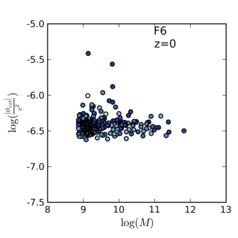} 
\\
\includegraphics[keepaspectratio,width=0.3\linewidth,height=0.14\textheight]{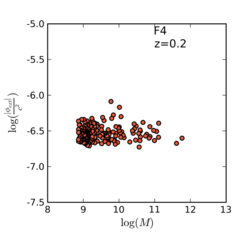} &
\includegraphics[keepaspectratio,width=0.3\linewidth,height=0.14\textheight]{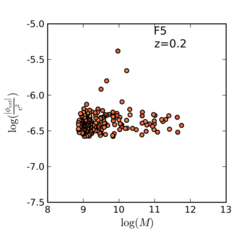} &
\includegraphics[keepaspectratio,width=0.3\linewidth,height=0.14\textheight]{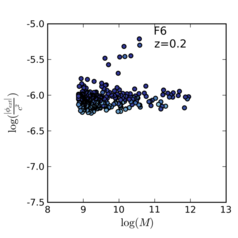} 
\\
\includegraphics[keepaspectratio,width=0.3\linewidth,height=0.14\textheight]{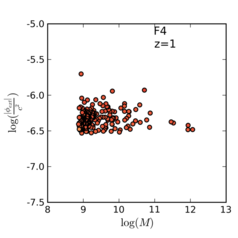} &
\includegraphics[keepaspectratio,width=0.3\linewidth,height=0.14\textheight]{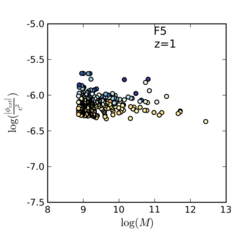} &
\includegraphics[keepaspectratio,width=0.3\linewidth,height=0.14\textheight]{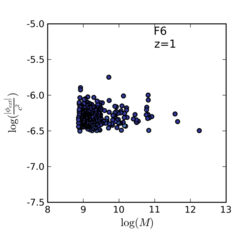} 
\\
\includegraphics[keepaspectratio,width=0.3\linewidth,height=0.14\textheight]{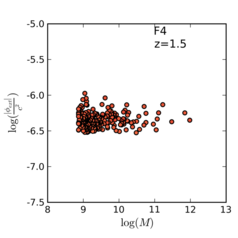} &
\includegraphics[keepaspectratio,width=0.3\linewidth,height=0.14\textheight]{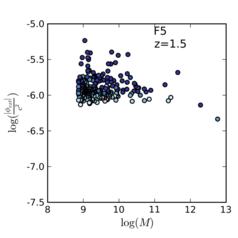} &
\includegraphics[keepaspectratio,width=0.3\linewidth,height=0.14\textheight]{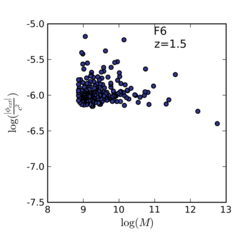} 
\\
\includegraphics[keepaspectratio,width=0.3\linewidth,height=0.14\textheight]{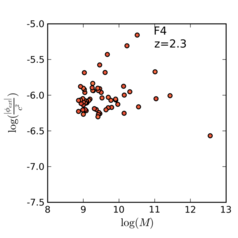} &
\includegraphics[keepaspectratio,width=0.3\linewidth,height=0.14\textheight]{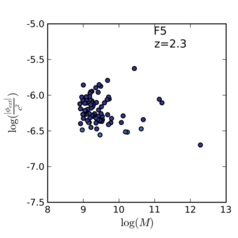} &
\includegraphics[keepaspectratio,width=0.3\linewidth,height=0.14\textheight]{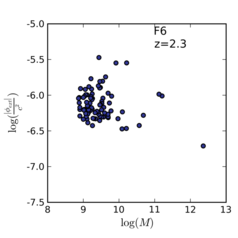} 
\\
\includegraphics[keepaspectratio,width=0.3\linewidth,height=0.14\textheight]{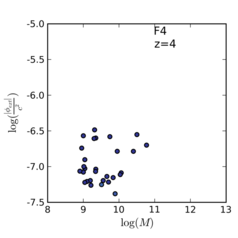} &
\includegraphics[keepaspectratio,width=0.3\linewidth,height=0.14\textheight]{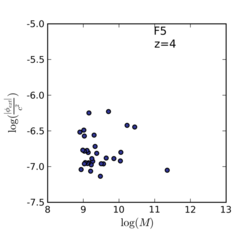} & 
\includegraphics[keepaspectratio,width=0.3\linewidth,height=0.14\textheight]{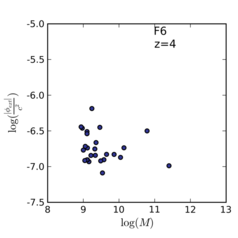} 
\end{array}
& \specialcell{\includegraphics[keepaspectratio,height=0.14\textheight]{{colorbar}.png}}
\end{array}$
\end{center}
\caption{$\phi_{\rm{ext}}$ as a function host halo mass, colored by $\Delta_M$. Here we see that $\phi_{\rm{ext}}$ is a reasonable proxy for screening for satellites.}
\label{phiext}
\end{figure*}

We note that all of our subhalo satellites were found to be too small to be self-screened. Because of this, any screening is due to the external field. The quantity $\log(|\phi_{ext}|)/c^2$ vs $\log(M_{dyn})$ is depicted in \textbf{Figure \ref{phiext}}. Focusing on the redshifts which show transitions we can see as in \cite{Cabre:2012cm} we could potentially make a cut in this plane to draw the distinction between screened and unscreened and that the point of this cut to reproduce the $\Delta_M$ results varies by redshift. There are clearly still secondary effects, but we show that $\phi_{ext}$ can serve as a reasonable proxy for screening of satellites. 

\begin{figure*}
\begin{center}$
\begin{array}{cc}
\begin{array}{ccc}
F4 & F5 & F6 \\    
\includegraphics[keepaspectratio,width=0.3\linewidth,height=0.14\textheight]{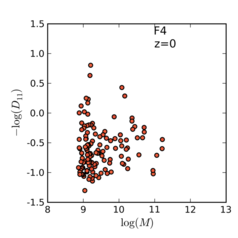} &
\includegraphics[keepaspectratio,width=0.3\linewidth,height=0.14\textheight]{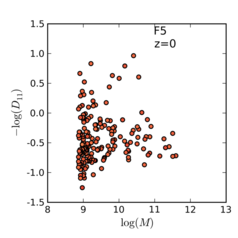} &
\includegraphics[keepaspectratio,width=0.3\linewidth,height=0.14\textheight]{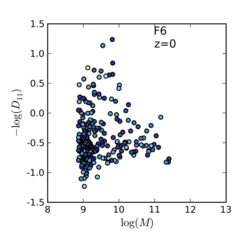} 
\\
\includegraphics[keepaspectratio,width=0.3\linewidth,height=0.14\textheight]{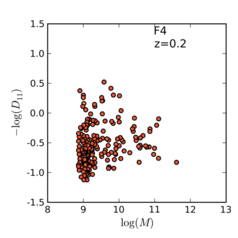} &
\includegraphics[keepaspectratio,width=0.3\linewidth,height=0.14\textheight]{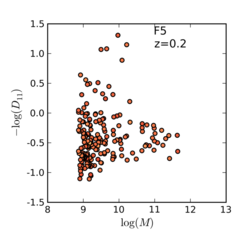} &
\includegraphics[keepaspectratio,width=0.3\linewidth,height=0.14\textheight]{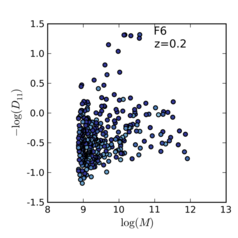} 
\\
\includegraphics[keepaspectratio,width=0.3\linewidth,height=0.14\textheight]{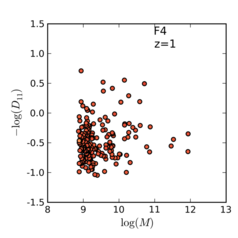} &
\includegraphics[keepaspectratio,width=0.3\linewidth,height=0.14\textheight]{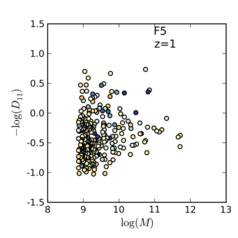} &
\includegraphics[keepaspectratio,width=0.3\linewidth,height=0.14\textheight]{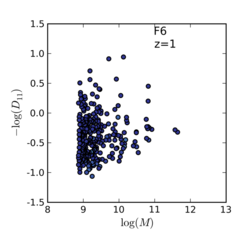} 
\\
\includegraphics[keepaspectratio,width=0.3\linewidth,height=0.14\textheight]{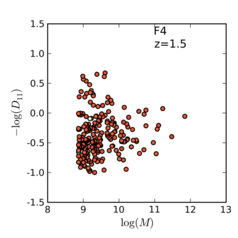} &
\includegraphics[keepaspectratio,width=0.3\linewidth,height=0.14\textheight]{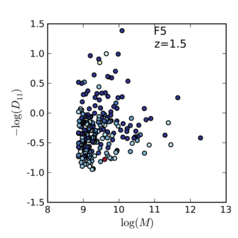} &
\includegraphics[keepaspectratio,width=0.3\linewidth,height=0.14\textheight]{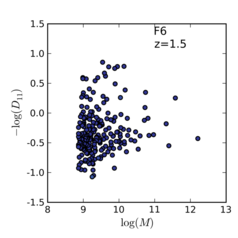} 
\\
\includegraphics[keepaspectratio,width=0.3\linewidth,height=0.14\textheight]{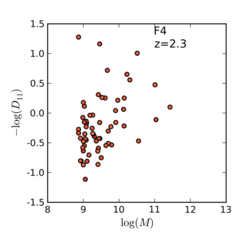} &
\includegraphics[keepaspectratio,width=0.3\linewidth,height=0.14\textheight]{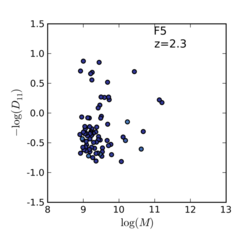} &
\includegraphics[keepaspectratio,width=0.3\linewidth,height=0.14\textheight]{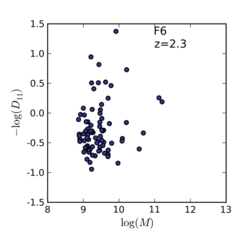}
\\
\includegraphics[keepaspectratio,width=0.3\linewidth,height=0.14\textheight]{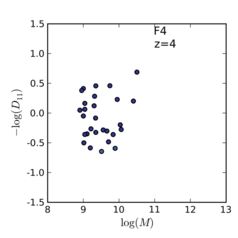} &
\includegraphics[keepaspectratio,width=0.3\linewidth,height=0.14\textheight]{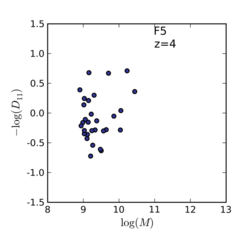} & 
\includegraphics[keepaspectratio,width=0.3\linewidth,height=0.14\textheight]{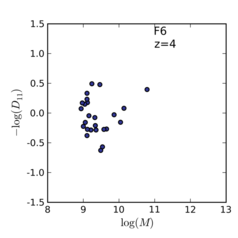} 
\end{array}
& \specialcell{\includegraphics[keepaspectratio,height=0.14\textheight]{{colorbar}.png}}
\end{array}$
\end{center}
\caption{$D_{11}$ as a function host halo mass, colored by $\Delta_M$. Here we see that $D_{11}$ is not a particularly good proxy for screening for satellites.}
\label{dfn}
\end{figure*}

We could also use the parameter $D_{fn}$ as a complementary proxy of screening. This parameter was introduced by \cite{Haas:2011mt} and deployed likewise by \cite{Cabre:2012cm}. For a given galaxy it is defined in terms of its neighbors:
\begin{equation}
    D_{fn} = d/r_{\rm{vir}}
\end{equation}
in this case,  $r_{\rm{vir}}$ is the virial radius of the nearest neighbor galaxy and $d$ is the distance to the $n$-th nearest neighbor with mass $f$ times higher than the galaxy considered. \cite{Haas:2011mt} show that the choice of $D_{11}$ is a good indicator of environmental screening. \cite{Cabre:2012cm} find a cut in this plane can be predictive of screening. \textbf{Figure \ref{dfn}} depicts the plot of $D_{11}$ vs $\log(M_{dyn})$ and coloring by the relative difference $\Delta_M$ between the dynamical and lensing masses of the same object. We show firstly that such a cut would need to vary by redshift and model, and that for satellites, such a cut is not straightforward and does not, focusing on the model F5 at $z=1$ and $z=1.5$, reproduce $\Delta_M$ from the simulation. Thus while $\phi_{\rm{ext}}$ remains an excellent proxy for screening for both halos and subhalos, $D_{11}$ seems to be applicable to halos only. It would be of observational interest to devise a $D_{11}$ analogue criterion applicable to satellites.

\begin{figure*}
\begin{center}$
 \begin{array}{ccc}
\includegraphics[keepaspectratio,width=0.3\linewidth,height=0.14\textheight]{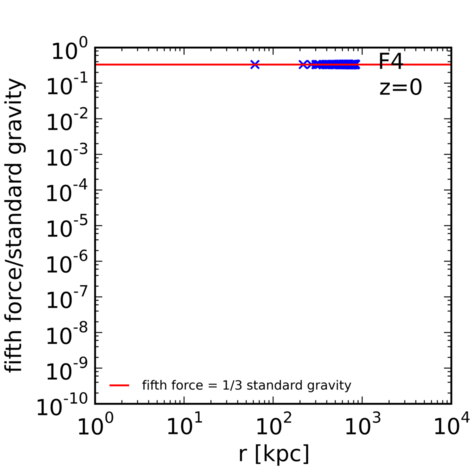} &
 \includegraphics[keepaspectratio,width=0.3\linewidth,height=0.14\textheight]{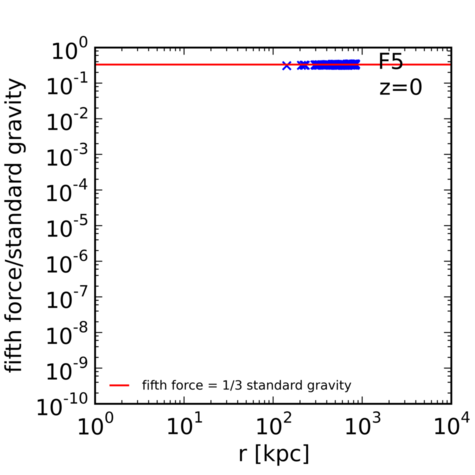} &
 \includegraphics[keepaspectratio,width=0.3\linewidth,height=0.14\textheight]{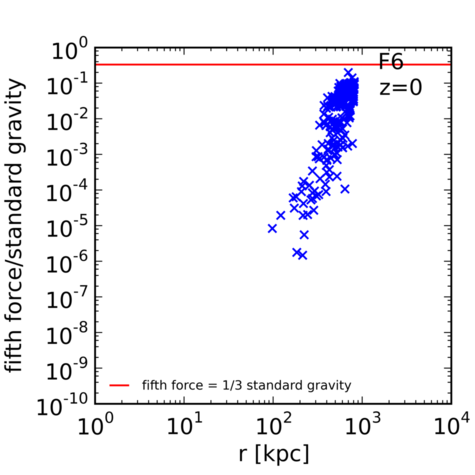}
 \\
 \includegraphics[keepaspectratio,width=0.3\linewidth,height=0.14\textheight]{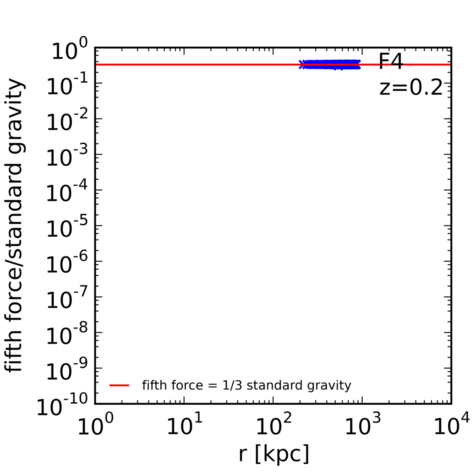} &
 \includegraphics[keepaspectratio,width=0.3\linewidth,height=0.14\textheight]{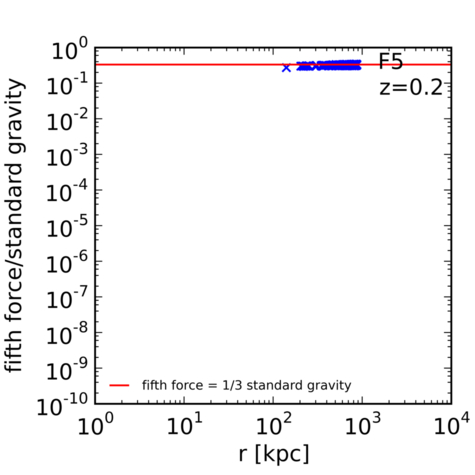} &
 \includegraphics[keepaspectratio,width=0.3\linewidth,height=0.14\textheight]{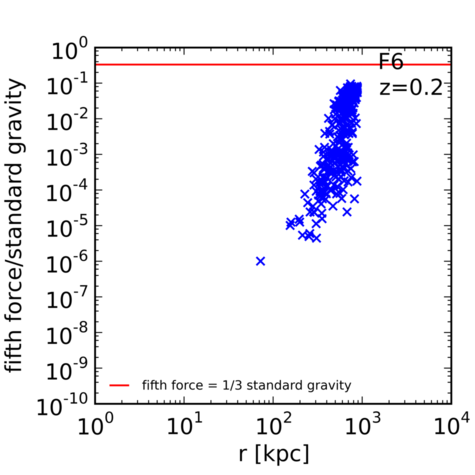}
 \\
 \includegraphics[keepaspectratio,width=0.3\linewidth,height=0.14\textheight]{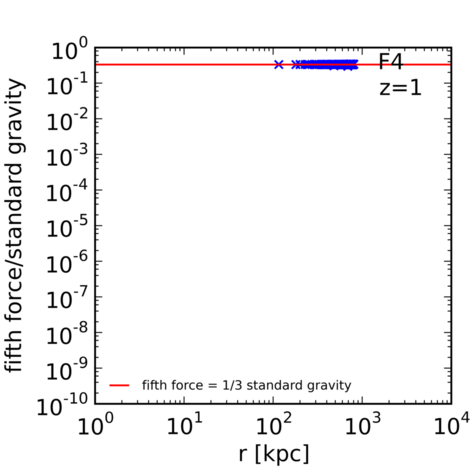} &
 \includegraphics[keepaspectratio,width=0.3\linewidth,height=0.14\textheight]{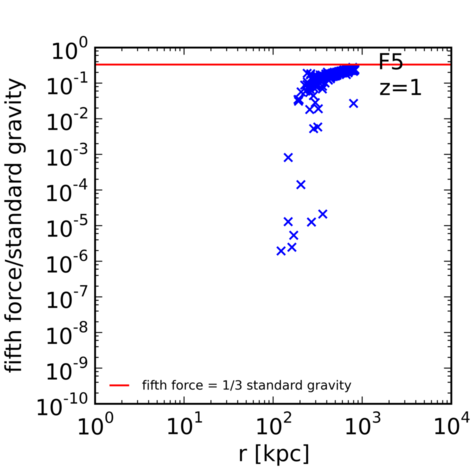} &
 \includegraphics[keepaspectratio,width=0.3\linewidth,height=0.14\textheight]{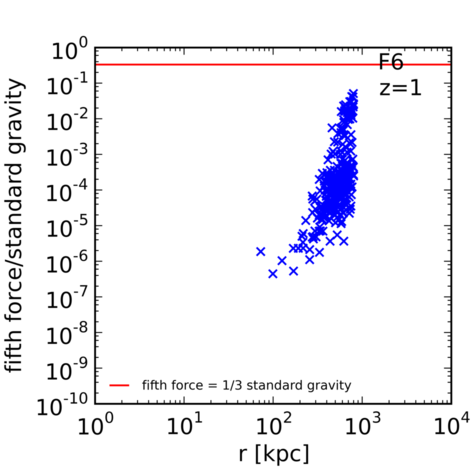}
 \\
 \includegraphics[keepaspectratio,width=0.3\linewidth,height=0.14\textheight]{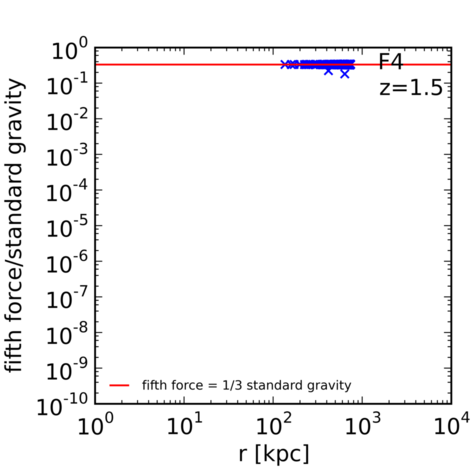} &
 \includegraphics[keepaspectratio,width=0.3\linewidth,height=0.14\textheight]{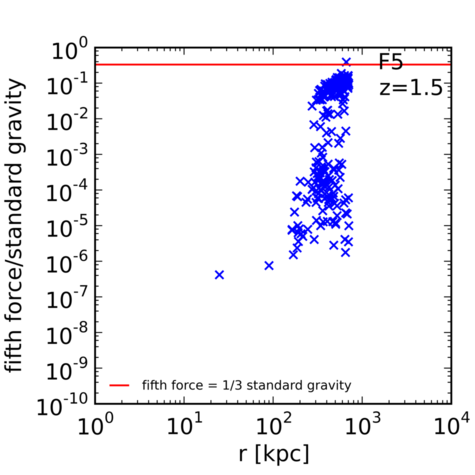} &
 \includegraphics[keepaspectratio,width=0.3\linewidth,height=0.14\textheight]{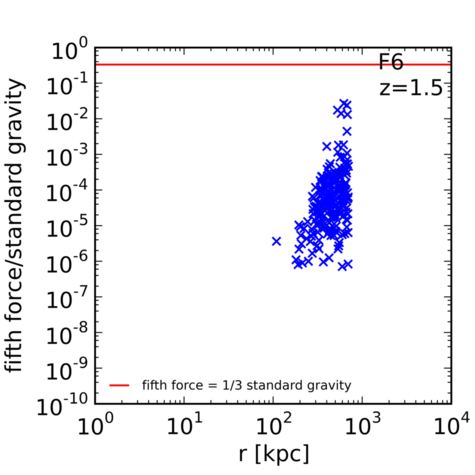}
 \\
 \includegraphics[keepaspectratio,width=0.3\linewidth,height=0.14\textheight]{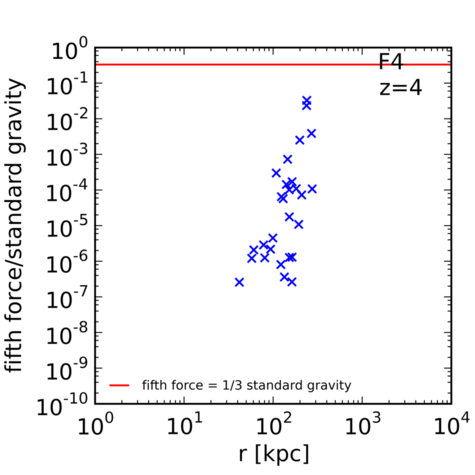} &
 \includegraphics[keepaspectratio,width=0.3\linewidth,height=0.14\textheight]{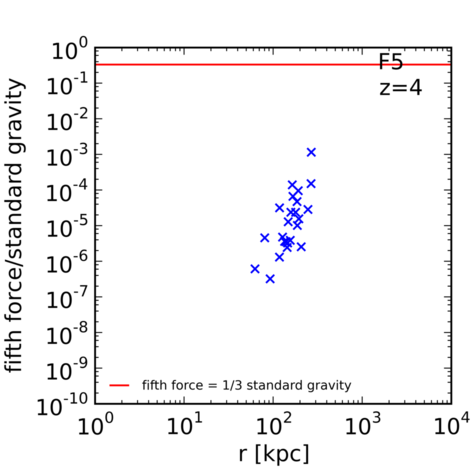} &
 \includegraphics[keepaspectratio,width=0.3\linewidth,height=0.14\textheight]{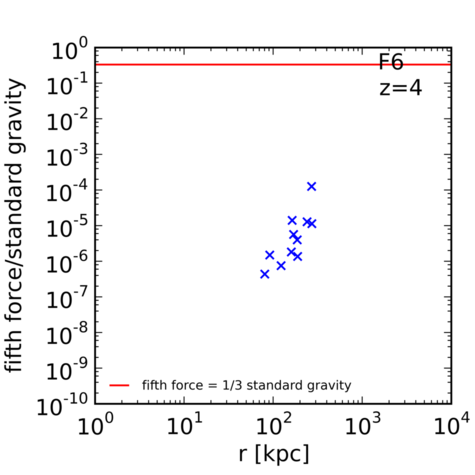}
 \end{array}$
\end{center}
\caption{Force ratio fifth for satellites versus radius. Comparing to \textbf{Figure \ref{ratioforces}} for satellites alone we see that indeed the modified gravity effects as measured by the satellite population well traces the result given by the underlying smooth dark matter density distribution.}
\label{ratioforcessats}
\end{figure*}

Next we show that the screening of satellites can be used as proxies for the results given by the underlying smooth dark matter distribution. Reproducing \textbf{Figure \ref{ratioforces}} for satellites alone we see in \textbf{Figure \ref{ratioforcessats}} that indeed the modified gravity effects as measured by the satellite population well traces the result given by the underlying smooth dark matter density distribution.

\begin{figure*}    
\begin{center}$
\begin{array}{cc}
\begin{array}{cccc}
z & F4 & F5 & F6 \\
0 & \includegraphics[keepaspectratio,width=0.3\linewidth,height=0.14\textheight]{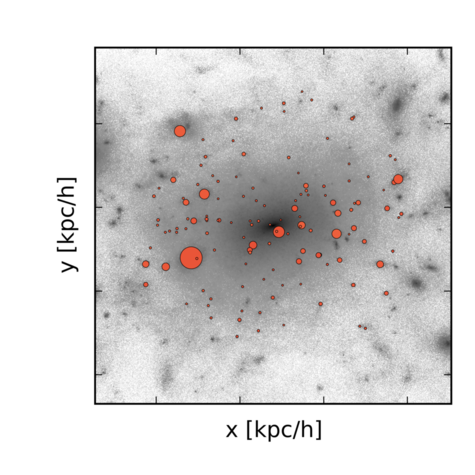} &
\includegraphics[keepaspectratio,width=0.3\linewidth,height=0.14\textheight]{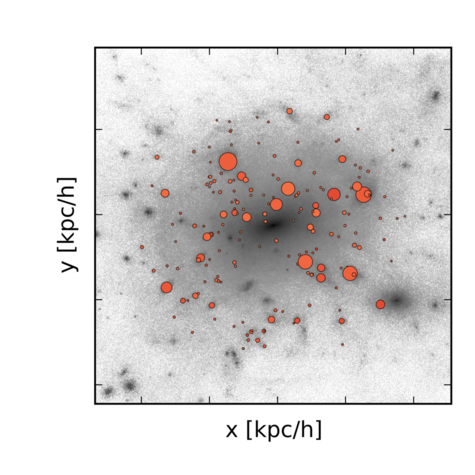} &
\includegraphics[keepaspectratio,width=0.3\linewidth,height=0.14\textheight]{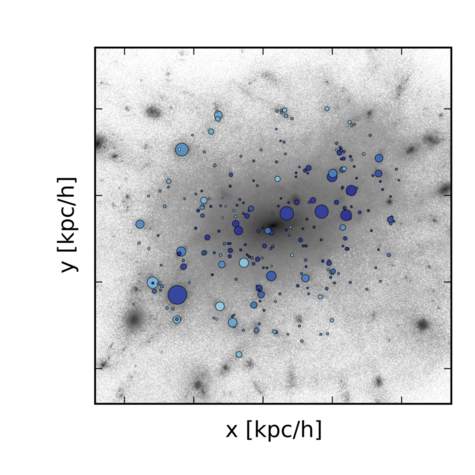}
\\
0.2 & \includegraphics[keepaspectratio,width=0.3\linewidth,height=0.14\textheight]{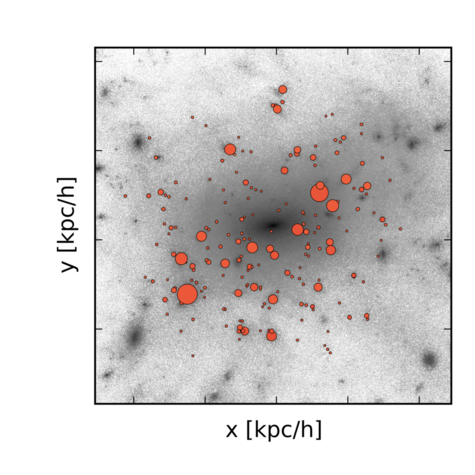} &
\includegraphics[keepaspectratio,width=0.3\linewidth,height=0.14\textheight]{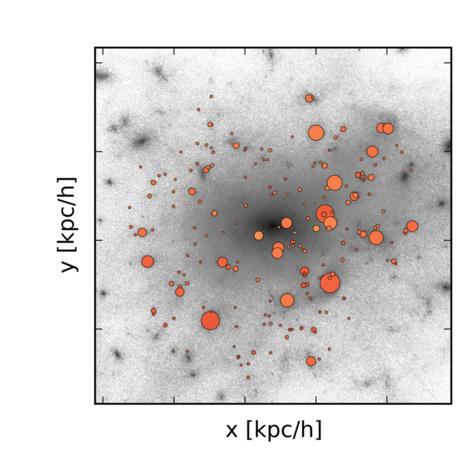} &
\includegraphics[keepaspectratio,width=0.3\linewidth,height=0.14\textheight]{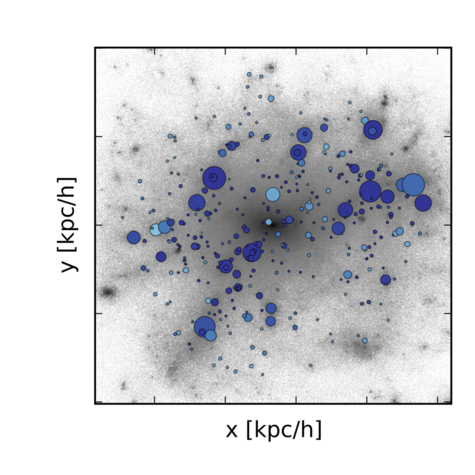}
\\
1 & \includegraphics[keepaspectratio,width=0.3\linewidth,height=0.14\textheight]{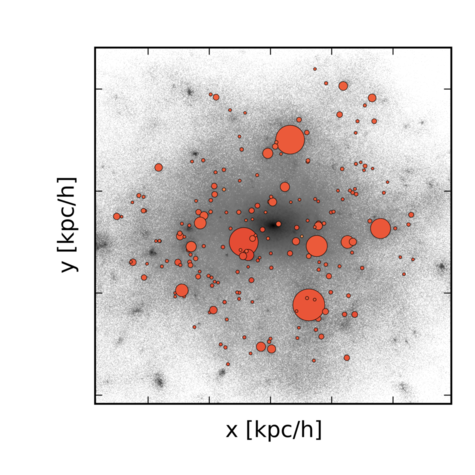} &
\includegraphics[keepaspectratio,width=0.3\linewidth,height=0.14\textheight]{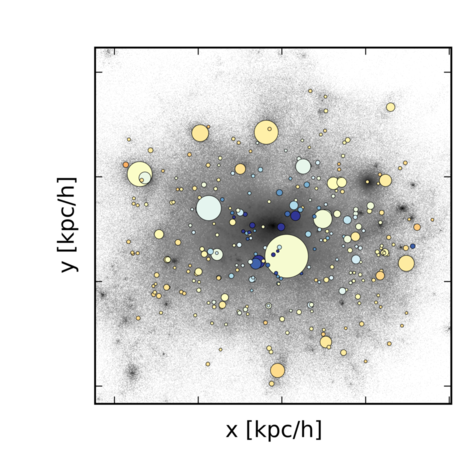} &
\includegraphics[keepaspectratio,width=0.3\linewidth,height=0.14\textheight]{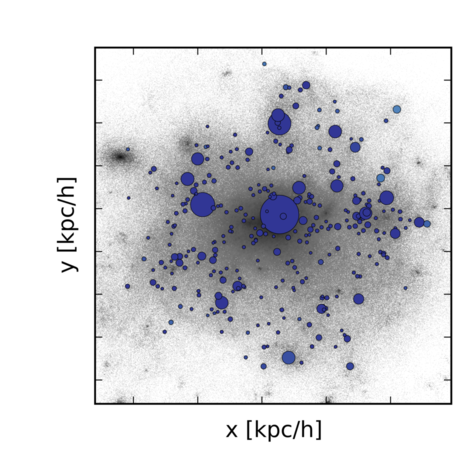}
\\
1.5 &\includegraphics[keepaspectratio,width=0.3\linewidth,height=0.14\textheight]{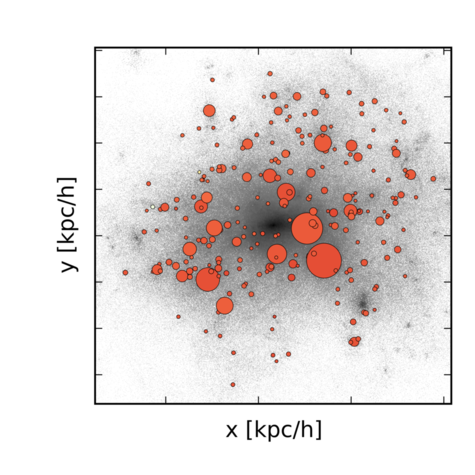} &
\includegraphics[keepaspectratio,width=0.3\linewidth,height=0.14\textheight]{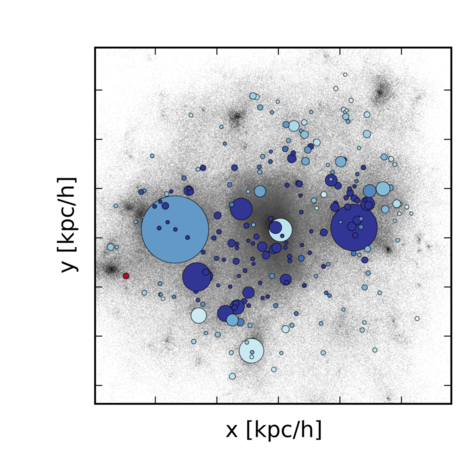} &
\includegraphics[keepaspectratio,width=0.3\linewidth,height=0.14\textheight]{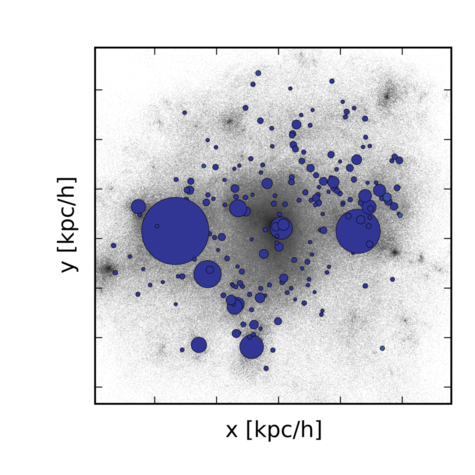}
\\
2.3 & \includegraphics[keepaspectratio,width=0.3\linewidth,height=0.14\textheight]{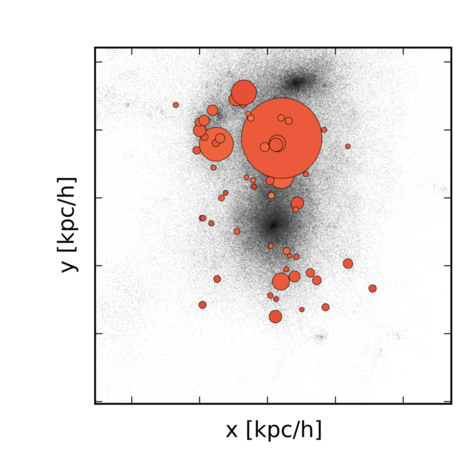} &
\includegraphics[keepaspectratio,width=0.3\linewidth,height=0.14\textheight]{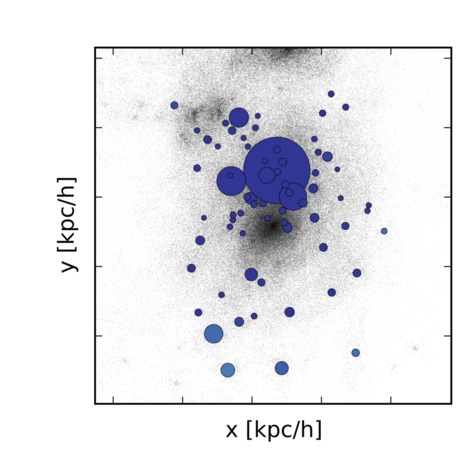} &
\includegraphics[keepaspectratio,width=0.3\linewidth,height=0.14\textheight]{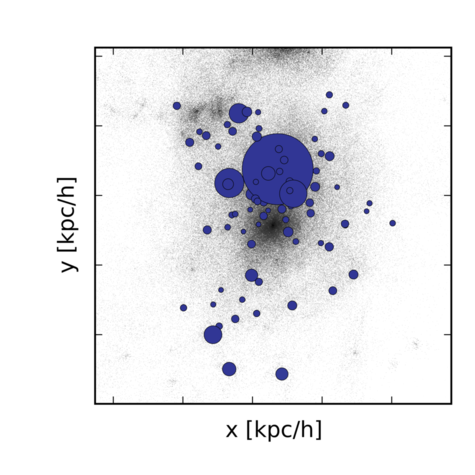}
\\
4 & \includegraphics[keepaspectratio,width=0.3\linewidth,height=0.14\textheight]{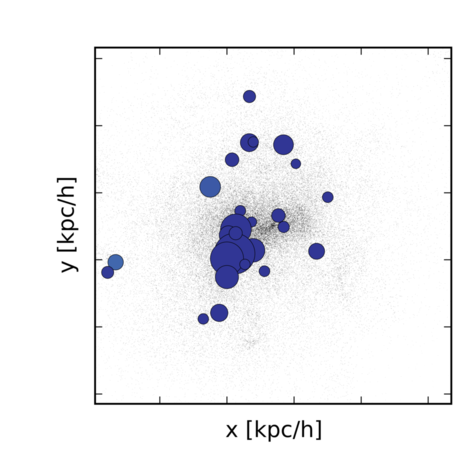} &
\includegraphics[keepaspectratio,width=0.3\linewidth,height=0.14\textheight]{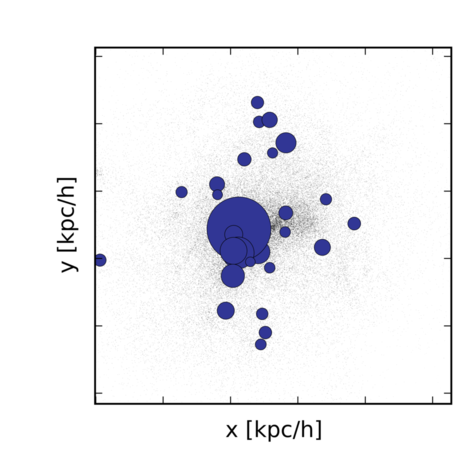} &
\includegraphics[keepaspectratio,width=0.3\linewidth,height=0.14\textheight]{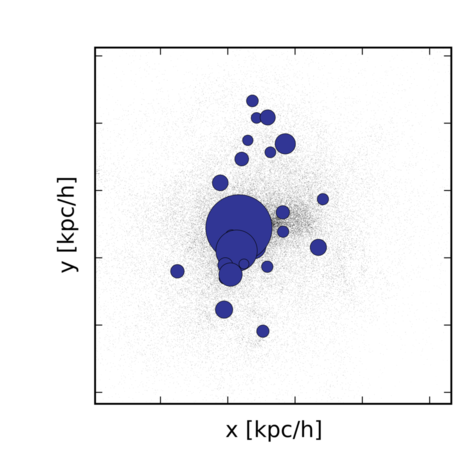}
\end{array}
& \specialcell{\includegraphics[keepaspectratio,height=0.14\textheight]{{colorbar}.png}}
\end{array}$
\end{center}
\caption{Satellites colored by $\Delta_M$. Scale of each image is 2$r_{\rm{vir}}$ at that redshift, and the size of each satellite is the size computed by \texttt{ROCKSTAR}. $\Delta_M$ colormap is corresponds to the identical values as in  \textbf{Figure \ref{deltam}}. A density map, depicting the line of sight density of the matter distribution, is presented in a greyscale color map in the background, to put the screening of the satellites in an environmental context.}
\label{satellites}
\end{figure*}

Finally, we present the maps in two dimensions for satellite galaxies in \textbf{Figure \ref{satellites}}, where the satellites are colored according to $\Delta_M$ with the same colormap as in \textbf{Figure \ref{deltam}}. This gives us insight into how subhalos are screened as a function of their position relative to the host halo. We see that the screening map is complex, and again focusing on the transition epochs of F5 $z=1.5$ and $z=1$ we see that while in general the outermost satellites become unscreened first, we can find counterexamples in the innermost subhalos of the $z=1$ epoch where a satellite becomes unscreened before any of its neighbor halos. Likewise, at the $z=1.5$ we can find subhalos in the interior already beginning to be unscreened while regions in the exterior continue to be fully screened. 

Thus, given that the weakest $f(R)$ gravity model F6 is beginning a transition epoch at present, observational signatures would be expected to be strongest at the outer regions of a massive halo, but could in principle exist as well in satellites closer to the interior of the main halo. 

We next explore the effect in three dimensions to see the environmental effects on $\Delta_M$ to not merely be a function of distance to the center of the main halo, focusing on the transition epoch of the F5 model, but noting that the behavior is expected to occur at higher redshift for stronger $f(R)$ models and at lower redshift for weaker $f(R)$ models.  For the F6 model this effect would be just at its onset in the present day. As in 2D, we again observe that in general, modified gravity affects the outskirts of the halo first and progresses inward.

\begin{figure*}
\begin{center}
 \includegraphics[keepaspectratio,width=\linewidth]{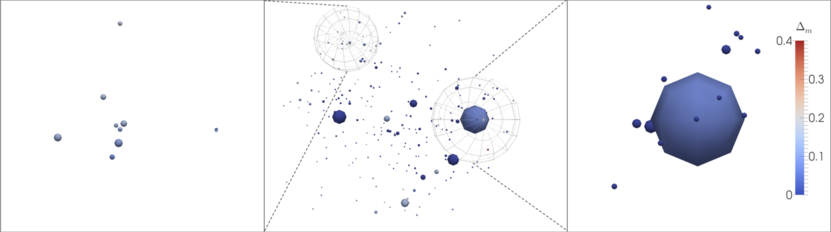}
 
\end{center}
\caption{Here we zoom in a radius of 200 kpc at a distance of $\approx 600$ kpc from the main halo both around the center of the most massive subhalo (right inset), and near no massive subhalos (left inset). The middle figures present the selections in context. Here we can see that the satellites around the most massive subhalo are entirely screened, while the satellites not in the vicinity of a massive subhalo all meet the criterion $\Delta_M>0.1$ and thus are beginning to be effected by enhanced gravitational force.}
\label{z1p5zoom}
\end{figure*}

In the middle inset of \textbf{Figure \ref{z1p5zoom}} we see that the satellites around the most massive subhalo are entirely screened, while the satellites far from a massive subhalo at the same distance from the main halo all meet the criterion $\Delta_M>0.1$ and thus are beginning to be effected by enhanced gravitational force.  We can see that the radial progression is roughly spherical but can be influenced for satellites near large subhalos, whose gravitational effect causes these satellites to remain screened and immune to the modified gravitational effect until higher redshift.  We additionally see that this behavior is self-similar, namely that the satellites within or nearby massive subhalos likewise remain screened longer in the inner part of the subhalo than the outer part, as is the case for subhalos within the main halo. Observing this effect in simulation for satellites is only possible due to the high resolution of our simulation. 

Further emphasizing the environmental effect we we zoom in a radius of 200 kpc both around the center of the most massive subhalo and in the field, near no massive subhalos. Both zooms are at the same distance, $\sim 600$kpc far from the main halo center. The right inset \textbf{Figure \ref{z1p5zoom}}  depicts a spherical region near a massive subhalo; all satellites in the vicinity are fully screened. The left inset \textbf{Figure \ref{z1p5zoom}} depicts a spherical region near no massive subhalos; all satellites are beginning the transition from screened to unscreened. 
     
     Observationally, this transition region could be probed by probing the underlying potential, the large dark matter subhalos will be detectable solely through their gravitational effects leaving velocity signatures on their stellar content. We predict, for a particular transition redshift to each model, objects residing in the outskirts to largely be unscreened, but objects lying within a large subhalo to be screened. These two populations could in principle be separated, and observationally compared. The unscreened population is expected to show signs of modified gravitational effects including systematically higher velocity dispersions. 
     
     Observing a lack of such signatures differentiating the two populations in a redshift range could rule out a particular $f(R)$ model parameter, and at all redshifts would rule out all but the weakest $f(R)$ models, in which this transition regime occurs in the far future. We predict this transition regime begins at $z\approx3$ for the F4 model, at $z\approx2$ for the F5 model and at $z\approx0$ for the F6 model. For models stronger than F4, not observationally favored, the transition would occur at higher redshift; for models weaker than F6 this transition regime would occur in the future, and thus be unobservable at present. Thus observations at $z\approx3$, $z\approx2$ and $z\approx0$ can be used to discriminate between the F4, F5 and F6 models or favor $\Lambda$CDM or equivalently a very weak $f(R)$ model.

\section{Conclusions} 
\label{sec:conclusions}
We build upon work by \cite{Li:2012ck} focusing on a particular modified gravity model, in which general relativity is modified within chameleon $f(R)$ gravity according to the Hu-Sawicki prescription, performing the highest resolution N-body zoom simulations in this model to date with the \texttt{ECOSMOG} code. We measure the surface density and velocity dispersion profiles and quantify their variation over chosen lines of sight, for select redshifts and model parameters the velocity dispersion profiles are robustly distinguishable from $\Lambda$CDM for our particular halo, with the caveat that this can be in general dependent on the merging history of the the halo under consideration. We quantify both the relative lensing signal profile, and the profile of $\Delta_M$ as a function of radius, relating an observable to the underlying physics of each model and showing that for each model, at a given redshift there is a characteristic radius from the main halo below which screening is fully effective. We show that this radius shrinks as time progresses in each model.

With the ability to analyze substructure in an N-body $f(R)$ simulation for the first time, we show that particles residing outside of halos are almost exclusively in the modified gravity regime. Building upon work by \cite{Cabre:2012cm} we show that the observable $|\phi_{\rm{ext}}|$ is a reasonable proxy for whether a satellite is subject to screening or not, but the neighbor based criterion $D_{11}$ deployed successfully for halos, is not as robust for satellites. Finally we show that even at the present day, in the weakest $f(R)$ model studied, a portion of halos will be subject to the unscreened fifth force. We show the transition epoch from the completely screened to completely unscreened regime for the main halo is given by the epoch at which the gravitational potential of the halo is on the order of the background value of $f_R$, $f_{\bar{R}z}$. Thus we expect there to be a range of halo screening values at various radii with in general the outermost radii being less screened than the innermost, and smaller halos being more likely to remain screened longer than larger halos.

Our results are as follows for a Virgo like halo:
\begin{itemize}
\item[(i)] We demonstrate, for a particular halo, that the velocity dispersion profiles of the F4 and F5 models could discriminate robustly from $\Lambda$CDM and the F6 model, and that F4 and F5 can be distinguished from each other at high redshift via this mechanism, or at intermediate redshift using the lensing signal profile. This distinguishability is expected to depend upon the merger history of the halo under consideration. (\textbf{Section 5, Section 6})
\item[(ii)] All models studied have a characteristic radius above which the chameleon mechanism fails to work at a given redshift. (\textbf{Section 7})
\item[(iii)] Particles not residing in halos in general are not screened by the chameleon affect across all models and redshifts. (\textbf{Section 7})
\item[(iv)] We quantify the environmental dependence of the screening effect on satellite galaxies and predict that in the weakest $f(R)$ gravity models, satellite galaxies at all radii could show the effects of transitioning into the modified gravity regime at the present day, with the strength of the effect increasing with radius on the whole. (\textbf{Section 8})
\item[(v)] For F6 we show that additional information, such as the behavior of satellites at the present day, could be used to discriminate this weaker model from $\Lambda$CDM, important as it is such weak models that are most favored but also the most difficult to probe observationally. (\textbf{Section 8})
\item[(vi)] We show that the observable screening of satellites can be used as reliable proxies for the results given by the underlying smooth dark matter distribution. (\textbf{Section 8})
\item[(vii)] The quantity $|\phi_{\rm{ext}}|$ is a reasonable proxy for the screening level of a satellite. (\textbf{Section 8})
\end{itemize}

\section{Acknowledgements}
Simulations were performed on the Monte Rosa system at the Swiss Supercomputing Center (CSCS). C.C.M. was supported by the HP2C program through the Swiss National Science Foundation. B. L. is supported by the Royal Astronomical Society and Durham University.

\clearpage
\bibliographystyle{mn2e}
\bibliography{FRMNRAS}

\end{document}